\pdfoutput=1
\documentclass{emulateapj}

\usepackage{graphicx}
\usepackage{times}
\usepackage{apjfonts}
\usepackage{amsmath}
\usepackage[raggedright]{subfigure}
\usepackage{color}
\newcommand{\rme}{e} 

\newcommand{\Eref}[1]{Equation~(\ref{#1})}
\newcommand{\dfBW}{\Delta f_{\textrm{\mbox{\tiny{BW}}}}}
\newcommand{\degrees}{\ifmmode^{\circ}\else$^{\circ}$\fi}
\newcommand{\amin}{\ifmmode^{\prime}\else$^{\prime}$\fi}
\newcommand{\asec}{\ifmmode^{\prime\prime}\else$^{\prime\prime}$\fi}
\def\EAH{{\it Einstein@Home}}

\DeclareGraphicsExtensions{.jpg}

\shorttitle{GAMMAY-RAY PULSARS VIA NEW SEARCH METHOD} 
\shortauthors{PLETSCH ET AL.}

\begin{document}

\title{Discovery of nine gamma-ray pulsars in \textit{FERMI}-LAT data using a new blind search method} 

\author{
H.~J.~Pletsch\altaffilmark{1,2,3},
L.~Guillemot\altaffilmark{4,5},
B.~Allen\altaffilmark{1,6,2},
M.~Kramer\altaffilmark{4,7},
C.~Aulbert\altaffilmark{1,2},
H.~Fehrmann\altaffilmark{1,2},
P.~S.~Ray\altaffilmark{8},
E.~D.~Barr\altaffilmark{4},
A.~Belfiore\altaffilmark{9,10,11},
F.~Camilo\altaffilmark{12},
P.~A.~Caraveo\altaffilmark{11},
\"O.~\c{C}elik\altaffilmark{13,14,15},
D.~J.~Champion\altaffilmark{4},
M.~Dormody\altaffilmark{9},
R.~P.~Eatough\altaffilmark{4},
E.~C.~Ferrara\altaffilmark{13},
P.~C.~C.~Freire\altaffilmark{4},
J.~W.~T.~Hessels\altaffilmark{16,17},
M.~Keith\altaffilmark{18},
M.~Kerr\altaffilmark{19},
A.~de~Luca\altaffilmark{20},
A.~G.~Lyne\altaffilmark{7},
M.~Marelli\altaffilmark{11},
M.~A.~McLaughlin\altaffilmark{21},
D.~Parent\altaffilmark{22},
S.~M.~Ransom\altaffilmark{23},
M.~Razzano\altaffilmark{9,24},
W.~Reich\altaffilmark{4},
P.~M.~Saz~Parkinson\altaffilmark{9},
B.~W.~Stappers\altaffilmark{7}, and
M.~T.~Wolff\altaffilmark{8}
}
\altaffiltext{1}{Albert-Einstein-Institut, Max-Planck-Institut f\"ur Gravitationsphysik, D-30167 Hannover, Germany}
\altaffiltext{2}{Institut f\"ur Gravitationsphysik, Leibniz Universit\"at Hannover, D-30167 Hannover, Germany}
\altaffiltext{3}{email: holger.pletsch@aei.mpg.de}
\altaffiltext{4}{Max-Planck-Institut f\"ur Radioastronomie, Auf dem H\"ugel 69, 53121 Bonn, Germany}
\altaffiltext{5}{email: guillemo@mpifr-bonn.mpg.de}
\altaffiltext{6}{Physics Department, University of Wisconsin -- Milwaukee, Milwaukee, WI 53211, USA}
\altaffiltext{7}{Jodrell Bank Centre for Astrophysics, School of Physics and Astronomy, The University of Manchester, M13 9PL, UK}
\altaffiltext{8}{Space Science Division, Naval Research Laboratory, Washington, DC 20375-5352, USA}
\altaffiltext{9}{Santa Cruz Institute for Particle Physics, Department of Physics and Department of Astronomy and Astrophysics, University of California at Santa Cruz, Santa Cruz, CA 95064, USA}
\altaffiltext{10}{Universit\`a degli Studi di Pavia, 27100 Pavia, Italy}
\altaffiltext{11}{INAF-Istituto di Astrofisica Spaziale e Fisica Cosmica, I-20133 Milano, Italy}
\altaffiltext{12}{Columbia Astrophysics Laboratory, Columbia University, New York, NY 10027, USA}
\altaffiltext{13}{NASA Goddard Space Flight Center, Greenbelt, MD 20771, USA}
\altaffiltext{14}{Center for Research and Exploration in Space Science and Technology (CRESST) and NASA Goddard Space Flight Center, Greenbelt, MD 20771, USA}
\altaffiltext{15}{Department of Physics and Center for Space Sciences and Technology, University of Maryland Baltimore County, Baltimore, MD 21250, USA}
\altaffiltext{16}{Astronomical Institute ``Anton Pannekoek,'' University of Amsterdam, Postbus 94249, 1090 GE Amsterdam, Netherlands}
\altaffiltext{17}{Netherlands Institute for Radio Astronomy (ASTRON), Postbus 2, 7990 AA Dwingeloo, Netherlands}
\altaffiltext{18}{CSIRO Astronomy and Space Science, Australia Telescope National Facility, Epping NSW 1710, Australia}
\altaffiltext{19}{W. W. Hansen Experimental Physics Laboratory, Kavli Institute for Particle Astrophysics and Cosmology, Department of Physics and SLAC National Accelerator Laboratory, Stanford University, Stanford, CA 94305, USA}
\altaffiltext{20}{Istituto Universitario di Studi Superiori (IUSS), I-27100 Pavia, Italy}
\altaffiltext{21}{Department of Physics, West Virginia University, Morgantown, WV 26506, USA}
\altaffiltext{22}{Center for Earth Observing and Space Research, College of Science, George Mason University, Fairfax, VA 22030, resident at Naval Research Laboratory, Washington, DC 20375, USA}
\altaffiltext{23}{National Radio Astronomy Observatory (NRAO), Charlottesville, VA 22903, USA}
\altaffiltext{24}{Istituto Nazionale di Fisica Nucleare, Sezione di Pisa, I-56127 Pisa, Italy}

\begin{abstract} 
\noindent
We report the discovery of nine previously unknown gamma-ray pulsars 
in a blind search of data from the \textit{Fermi} Large Area Telescope~(LAT).
The pulsars were found with a novel hierarchical search
method originally developed for detecting continuous gravitational
waves from rapidly rotating neutron stars.  Designed to find isolated
pulsars spinning at up to kHz frequencies, the new method is
computationally efficient, and incorporates several advances,
including a metric-based gridding of the search parameter space
(frequency, frequency derivative and sky location) and the use of
photon probability weights.
The nine pulsars have spin frequencies between
3~and~12~Hz, and characteristic ages ranging from~17~kyr 
to 3~Myr.  Two of them, \mbox{PSRs J1803--2149} and J2111+4606, 
are young and energetic Galactic-plane pulsars (spin-down power
above \mbox{6$\times$10$^{35}$~erg s$^{-1}$} and ages below~100~kyr).
The seven remaining pulsars, \mbox{PSRs J0106+4855}, \mbox{J0622+3749}, 
\mbox{J1620--4927}, \mbox{J1746--3239}, \mbox{J2028+3332}, \mbox{J2030+4415}, 
\mbox{J2139+4716}, are older and less energetic; 
two of them are located at higher Galactic latitudes \mbox{($|b| > 10\degr$)}.
PSR~J0106+4855 has the largest characteristic age 
(3~Myr) and the smallest surface magnetic field (2$\times$10$^{11}$G) 
of all LAT blind-search pulsars. PSR~J2139+4716 has the lowest 
spin-down power (\mbox{3$\times$10$^{33}$~erg s$^{-1}$})
among all non-recycled gamma-ray pulsars ever found.
Despite extensive multi-frequency observations, only PSR J0106+4855 has
detectable pulsations in the radio band.  The other eight pulsars
belong to the increasing population of radio-quiet gamma-ray
pulsars.
\end{abstract}


\section{Introduction}
\label{s:intro}

The \textit{Fermi Gamma-ray Space Telescope} has been operating since
launch in 2008 June.  The Large Area Telescope (LAT) on board the
\textit{Fermi} satellite has an effective area of \mbox{$\sim$ 0.8~m$^2$} (on-axis,
above 1~GeV) and is sensitive to photons with energies from 20~MeV to
more than 300~GeV \citep{generalfermilatref}. Post-processing assigns
an arrival time, energy~$E$, and direction to the photons; we call
these ``events''.  The arrival times are accurate to better than 1
$\mu$s, and the energies are accurate to within 15\% between 0.1 
and 10~GeV on-axis.  The directional
precision is energy-dependent: 68\% of photons have angular offset
less than $\sim$ 0$\fdg$8 $\times (E /\mathrm{GeV})^{-0.8}$ from the
true direction \citep{fermionorbitcalib}.

Gamma-ray pulsars are among the most interesting sources observed by
the \textit{Fermi} LAT.  These are rapidly-spinning neutron stars whose
regular beam of gamma-ray emissions passes by the detector with each
rotation.
The \textit{Fermi} LAT has detected gamma-ray pulsations from more than 
50 normal and millisecond pulsars (MSPs) previously discovered in 
radio-frequency searches  \citep[see
  e.g.;][]{FermiPSRCatalog,Ransom2011,Keith2011,Cognard2011,Theureau2011}.

In contrast, so-called ``blind'' searches for gamma-ray pulsars are
not guided by any prior knowledge (e.g. from radio or X-ray
observations) of the pulsars' parameters.  Previous blind
searches of the data recorded by the \textit{Fermi}-LAT have been
spectacularly successful; they have discovered 26 gamma-ray pulsars
\citep{16gammapuls2009,8gammapuls2010,2gammapuls2011}. This paper
describes the ``blind'' discovery of 9 additional pulsars using a new
method.

The blind search problem is computationally demanding because the
relevant pulsar parameters (typically sky-position, frequency $f$, and
spin-down rate $\dot f$) are not known a priori and must be explicitly
searched \citep{Chandler2001}. So far, blind searches of LAT data have 
used a clever ``time-differencing technique'' as described in
\citet{TimeDiffTech2006,Ziegler2008}.  
One powerful motivation for seeking even better methods is the 
application to MSPs, of which previous blind searches have not found any. 
The blind search for MSPs is more computationally challenging due to the 
higher frequency range that must be covered. Moreover, most MSPs are in 
binary systems, where the orbital modulation parameters must also be
searched, adding orders-of-magnitude to the computational
complexity and challenge.

This paper presents results from a new effort to find isolated
gamma-ray pulsars (including MSPs), using a novel method inspired by
computationally-efficient techniques recently developed to search
gravitational-wave detector data for weak continuous-wave signals from
rapidly-spinning isolated neutron stars
\citep[][]{PletschAllen,pletsch:scmetric}.  In particular, the search
method uses a recently-developed optimal incoherent-combination method
first described in~\citet{PletschAllen} together with a ``sliding
coherence window technique''~\citep{PletschSLCW2011}.

The method was originally intended to find isolated MSPs up to
1.4~kHz spin frequency with spin-down rates in the range 
\mbox{$-$5$\times$10$^{-13}$ Hz s$^{-1} \leq \dot f \leq 0$}, 
with a corresponding range of characteristic ages $\tau =-f/2\dot f$.
However, the search technique is also sensitive to normal (non-MSP)
isolated pulsars, and several normal pulsars were discovered
soon after we began.  This prompted us to extend the spin-down-rate
search range, in order to include younger objects
(having smaller~$\tau$), down to ages ($\sim$~kyr) comparable to
the Crab pulsar.  Here we report on the discovery of nine pulsars from 
this ongoing effort in which a total of 109 sources selected
from the \textit{Fermi}-LAT Second Source Catalog
\citep[][]{FermiSecondSourceCatalog} are being searched 
for new gamma-ray pulsars.

The outline of this paper is as follows.
Section~\ref{s:dataselection} describes the LAT data preparation, and
the selection of unidentified sources to search for previously unknown
gamma-ray pulsars.  The new hierarchical search method is explained in
Section~\ref{s:method} and illustrated in Section~\ref{s:example} with a
detailed example: our first discovery, \mbox{PSR~J1620--4927}.  
Section~\ref{s:allothers} presents the results for all the new
gamma-ray pulsars.  The search for counterparts in other regions of
the electromagnetic spectrum is discussed in
Section~\ref{s:counterparts}.  This is followed by a brief conclusion.

\section{Source selection and data preparation}
\label{s:dataselection}

The \textit{Fermi}-LAT Second Source Catalog
\citep[2FGL,][]{FermiSecondSourceCatalog} lists 1873 sources, 
described by fits to elliptically-shaped 95\%-confidence sky regions. 
Among these sources 576 are not associated with counterparts observed 
at other wavelengths and thus might contain unknown gamma-ray pulsars.

In searching for new gamma-ray pulsars, it is important to identify and exclude sources that are blazars.
Previously-observed \textit{Fermi}-LAT gamma-ray pulsars \citep[see
  e.g.][]{FermiPSRCatalog} have sharp cutoffs in their emission
spectra at a few GeV, and stable gamma-ray fluxes. In contrast, blazars
emit above 10~GeV, and their fluxes vary with time.  An illustration
of the different spectral properties and variability behavior can be
seen in Figure~17 of \citet{FermiSecondSourceCatalog}. Gamma-ray pulsars tend to 
have large curvature significances (``Signif\_Curve'' in the 2FGL catalog, 
which gives the improvement in the quality of the spectral fit when changing from 
a power law to a curved spectral model) and small 
variability indices (``Variability\_Index'' in the 2FGL catalog, a measure of flux 
instability over time). In contrast, blazars tend to have small curvature significances 
and large variability indices. 

For the search for gamma-ray pulsations 
we select 2FGL sources with curvature-significance values greater 
than 4$\sigma$ (here and throughout the manuscript, $\sigma$ denotes the 
standard deviation for a Gaussian distribution), and variability indices smaller than~41.6. 
We further select  bright objects by choosing sources with detection significances 
(``Signif\_Avg'') greater than 10$\sigma$. Finally, we restrict the list 
to sources with no known associations (``unassociated sources'') or associated 
with known supernova remnants (SNRs). By applying the above selection criteria 
to a preliminary version of the 2FGL catalog, we obtained a list of~109 2FGL 
sources to search for gamma-ray pulsations. (One source, 2FGL J0621.9+3750 was 
associated with an active galactic nucleus (AGN) in the final 2FGL catalog.)

The \textit{Fermi} Science Tools (ST)\footnote{See
  \url{http://fermi.gsfc.nasa.gov/ssc/data/analysis/scitools/overview.html}
  and
  \url{http://fermi.gsfc.nasa.gov/ssc/data/analysis/documentation/Cicerone/}
  for details and tutorials.} v9r23p1 are employed to select the
\textit{Fermi}-LAT events for our search.  Using
\textit{gtselect}, we take events from 2008 August~4 to 2011 April~6,
with reconstructed directions within 8$^\circ$ of the gamma-ray
sources, energies above 100~MeV, and zenith angles $\leq$ 100$^\circ$. 
Only events belonging to the Pass 6 ``Diffuse'' class are retained,
as those events have the highest probability of being photons
\citep{generalfermilatref}. 
For the event probability weighting described below 
we use the P6\_V11 Instrument Response Functions (IRFs). 
Using the \textit{gtmktime} tool, times
when the rocking angle of the satellite exceeded 52$^\circ$ are
excluded. We also require that DATA\_QUAL and LAT\_CONFIG are
unity, and that the Earth's limb does not impinge
upon the Region Of Interest~(ROI).  

As demonstrated in previous work \citep{Bickel2008,KerrWeightedH2011},
the sensitivity of gamma-ray pulsation searches can be improved by
weighting the $j$th event with a probability $w_j \in [0,1]$ that it
originated from a putative pulsar.  The photon weights $w_j$ are
calculated by using a full spectral model of the region around the
gamma-ray source, and by exploiting the IRFs to provide background
rejection which is superior to simple angular and energy
data-selection cuts.  For the first time in this paper such a
photon-weighting scheme has been applied in a blind search; further
details are given in Section~\ref{s:method}.

To calculate the weights $w_j$ for each event from a given source, 
we perform likelihood spectral analyses using the \textit{pointlike} tool
\citep[see][for a description]{KerrThesis}. For each selected source, 
we construct a spectral model for the region by including all sources 
of the 2FGL catalog 
found within 8$^\circ$ of the selected source, using the spectral forms 
given in the catalog. The spectra of the selected sources are modeled as 
exponentially cut-off power laws, typical of known gamma-ray pulsars, 
of the form $N_0 \left( E / \mathrm{GeV} \right)^{-\Gamma} \exp \left( - E / E_c \right)$, 
where $N_0$ is a normalization factor, $\Gamma$ is the photon index 
and $E_c$ is the cut-off energy. The Galactic and extragalactic diffuse emission 
and residual instrument background also enter the calculation of the weights.
The Galactic diffuse emission is modeled using
the \textit{gll\_iem\_v02\_P6\_V11\_DIFFUSE} map cube, while the extragalactic diffuse
and residual instrument backgrounds is modeled using the
\textit{isotropic\_iem\_v02\_P6\_V11\_DIFFUSE} template (a detailed description of these
background models can be found in Section~3 of \citet{FermiFirstSourceCatalog}). 
These models are available for download at the \textit{Fermi} Science Support
Center\footnote{http://fermi.gsfc.nasa.gov/ssc/data/access/lat/BackgroundModels.html}.
The tool \textit{gtsrcprob} is then used to calculate
the event weights $w_j$ based on the best-fit
spectral models obtained from the maximum likelihood analyses.

\section{The new search method}
\label{s:method}

In a year, the LAT detects of order 10$^3$ photons from a typical
gamma-ray pulsar; in the same year, a typical pulsar rotates at least
10$^8$ times around its axis.  The blind-search problem is to find a rotational-phase
model $\Phi(t) = 2 \pi (f t + {\dot f} t^2 / 2)$ and a sky-position that
match the Solar System Barycenter~(SSB) arrival times $t$ of the
different photons, where $\Phi$ denotes the rotation angle of the star
about its axis, in radians, measured from its starting position at
$t=0$, and observed at the SSB. The signal hypothesis is that the
photons' arrival times are ``clustered'' near specific ``orientations''
of the star (i.e., \mbox{$\Phi(t) \mod 2\pi$} deviates from uniformity on the 
interval~$[0,2\pi]$). The null hypothesis is that the arrival times of the photons 
are a random Poisson process.  In this paper we do not explicitly indicate
the dependence of $\Phi$ on $f$, $\dot f$ and sky position, but this dependence is
important and implicit in many formulae below.

To find a matching phase-model, a grid of ``templates'' in the
four-dimensional parameter space of sky position and $(f, \dot f)$ is
constructed.  Note that the 2FGL catalog sky positions of the targeted
unassociated sources based on the spatial distribution of events are
typically not precise enough for pulsar searches.  A search grid of
sky points around this catalog position is needed to reduce signal
loss arising from imperfect correction of the Doppler shifts caused by
the Earth's orbital motion around the SSB.  The need for sky gridding
is particularly acute for MSP spin frequencies.  Therefore, in
contrast to previously published blind searches\footnote{
  \citet{Ziegler2008} argues correctly that for two-week
  data-stretches, sky gridding is not essential.  However for data
  stretches of length comparable to a year or longer, sky gridding
  \emph{is} necessary to avoid significant loss of signal-to-noise ratio.},
we grid a circular sky-region centered on the 2FGL catalog source location 
using a radius which is 20\% larger than the semi-major axis of 
the 95\% confidence elliptical error region (given by the ``Conf\_95\_SemiMajor'' parameter).

Unfortunately the number of templates (grid points) required to
discretely cover the entire four-dimensional search parameter space
increases as a high power of the coherent integration-time. Hence a
fully-coherent approach for several years of data is computationally
impossible. Therefore, we employ a search strategy which is designed to 
achieve maximum overall sensitivity at fixed computing cost\footnote{If this constraint 
is removed, then it is obvious that a more sensitive method exists.}.  

To efficiently scan through years of \textit{Fermi}-LAT data for
previously unknown gamma-ray pulsars, we use a so-called hierarchical
search approach. This is analogous to hierarchical methods used in
searches for gravitational-wave pulsars
\citep{schutzpapa:1999,papa:2000wg,bc2:2000,S4EAH,S5R1EAH,cutler:2005}.
In a first semi-coherent stage, we here adopt the optimal metric-based
gridding methods described in \citet{PletschAllen} along with the
sliding coherence window technique \citep{PletschSLCW2011}.  In a
second stage, significant semi-coherent candidates are automatically
followed up in a fully-coherent analysis. Finally, a third stage
further refines coherent pulsar candidates by using higher harmonics.
Full details of the complete search scheme will be presented in
forthcoming work \citep{PletschGuillemot}.

Here we first describe the principle of the method, to firmly
establish the analogy with the existing gravitational-wave literature.
Then we describe what is done in practice, which is mathematically
equivalent (up to justifiable approximations) but computationally more
efficient.

In the first stage, a semi-coherent detection statistic $S$ is computed
for each template.  We refer to $S$ as ``semi-coherent''
because it is effectively the incoherent sum over several years, of terms
which are coherent over several days.  The coherent terms are the
power in Fourier bins, calculated by treating each photon arrival as a
delta function in time. 

Denoting the arrival time of the $j$th event (photon) at the SSB by $t_j$,
the coherent power~$P_\tau$ in a (Gaussian) window centered at time $\tau$ 
is defined by
\begin{equation}
   P_\tau = \biggl| \sum_{j=1}^N w_j \; \rme^{-i \Phi(t_j)} \rme^{-2
     \pi (t_j - \tau)^2/T^2}\biggr|^2.
   \label{e:Pdefinition}
\end{equation}
The sum is taken over all photons (here, $N=8000$) in the data set;
the effective window duration is $\int e^{-2 \pi \tau^2/T^2} d\tau
= T/\sqrt{2}$.  As described in Section~\ref{s:dataselection}, the
weights~$w_j$ estimate the probability that the photon comes from the
selected source.

To form the semi-coherent detection statistic~$S$, the values of
$P_\tau$ are summed (``incoherently combined'')
\begin{equation}
    S = \frac{2}{T} \int d\tau \; P_\tau - \sum_{j=1}^N w_j^2.
      \label{e:semidefinition}
\end{equation}
 Note that in this definition we have subtracted a constant
 (phase model indepdendent) term. Because it contains a
 Gaussian window, the integrand in \Eref{e:semidefinition} falls off
 exponentially at early and late time (large values of $|\tau|)$.
 Thus the limits of integration can be taken as the entire real line
 $\tau \in (-\infty, \infty)$; to good approximation this gives the
 same value as integrating only over the total observation interval
 (about 975 days in this search).

The semi-coherent detection statistic $S$ is an incoherent sum of
powers, which discards the phase information over time periods longer
than of order $T$.  This uniform overlap maximizes the search
sensitivity for fixed~$T$ and for fixed computational resources
\citep[][]{PletschSLCW2011}.  For computational efficiency, in this
search we choose the $N=8000$ photons with the highest probabilities
(largest values of $w_j$).

To understand how to compute $S$ efficiently, one can explicitly
evaluate \Eref{e:semidefinition}.  Completing the squares in the
product of the Gaussians and carrying out the integration over $\tau$,
one obtains
\begin{equation}
    S = \sum_{j=1}^N \sum_{k=1}^N w_j w_k {\rme}^{-i [\Phi(t_j) -
        \Phi(t_k)]} {\rme}^{-\pi (t_j - t_k)^2/T^2} - \sum_{j=1}^N w_j^2.
      \label{e:semidefinition2}
\end{equation}
Here the effective duration of the Gaussian window is $\int e^{-\pi
  \tau^2/T^2} d\tau = T$.  In practice, to compute the semi-coherent
power efficiently, we replace the Gaussian window in
\Eref{e:semidefinition2} with a rectangular window of the same
duration $T$, as given below in \Eref{e:semidefinition3}. In this
search, the width of the rectangular window is $T=2^{19}$~s ($\approx$
6~days).

The template grid in parameter space is the Cartesian product of a rectangular
two-dimensional grid in $(f, \dot f)$ and a sky grid which has
constant density when orthogonally projected onto the ecliptic
plane. The problem of constructing efficient search grids has been intensively 
studied in the context of gravitational wave searches
\citep[see e.g.][]{bccs1:1998,bc2:2000,prix:2007ks,PletschAllen,pletsch:scmetric}
and we employ these concepts here. The values of frequency are equally spaced, 
separated by the FFT frequency-bin width $\Delta f = 1/T$.  In practice, to reduce the 
fractional loss in~$S$ for frequencies not coinciding with Fourier frequencies, we use a
computationally-efficient interpolation, referred to as ``interbinning'' \citep{Ransom2002}.
The spacing in the other three dimensions is determined by a metric which measures
the fractional loss in the expected value of $S$, that arises if the
signal is not located exactly at a grid point \citep{Sathy1:1996,owen:1996me,prix:2007ks}. 
In spin-down~$\dot f$, we use a uniform grid spacing
$\Delta {\dot f} = {\sqrt{720 m}}/{\pi \gamma T^2}$. In this search $m=0.3$ is
the maximum tolerable fractional loss in $S$, and from
\cite{PletschAllen},
\begin{equation}
     \gamma^2 = 1 + \frac{60}{N} \sum_{j=1}^N \frac{\left( t_j-\bar t \right)^2}{T^2}, 
\end{equation}
where $\bar t = \sum_j t_j/N$ is the mean photon arrival time.  The
description of this grid can be found in \cite{PletschAllen} with a
detailed derivation in \cite{pletsch:scmetric}.  The grid in the sky
is determined by the same metric, permitting a maximum fractional loss
$m$ in the value of $S$.  
The spacing of the sky grid is determined by
the Doppler-shift arising from the Earth's (more precisely, the 
\textit{Fermi} satellite's) motion around the Sun.  At the north
Ecliptic pole the angular spacing is 
$\Delta \theta = {\sqrt{2m}}\,c/{\pi f D}$, where $c$ is the speed of light and $D$ is a
baseline distance (defined below). When the entire sky grid is
projected into the plane of the ecliptic, the grid-points are
uniformly spaced on the plane \citep{Astone,S4EAH,S5R1EAH}.  This
angular spacing is similar to the angular spacing in the diffraction
pattern of a two-slit system, where the wavelength is $c/f$ and the
separation of the two slits is the straight-line distance $D$ between
two points on the Earth's orbit about the Sun.  If the coherent
integration time $T$ is \emph{less} than half a year, then 
$D = (998\,\textrm{s}) \, c \, \sin (\pi T/1 \,\textrm{yr})$.  
If the coherent integration time $T$ is \emph{greater}
than half a year, then $D = (998\,\textrm{s}) \,c$ is the diameter of the Earth's orbit
about the Sun.

\begin{figure*}[t!!]
 	\centering
	\subfigure[\label{f:semicohJ1620}]
       	{\includegraphics[width=\columnwidth]{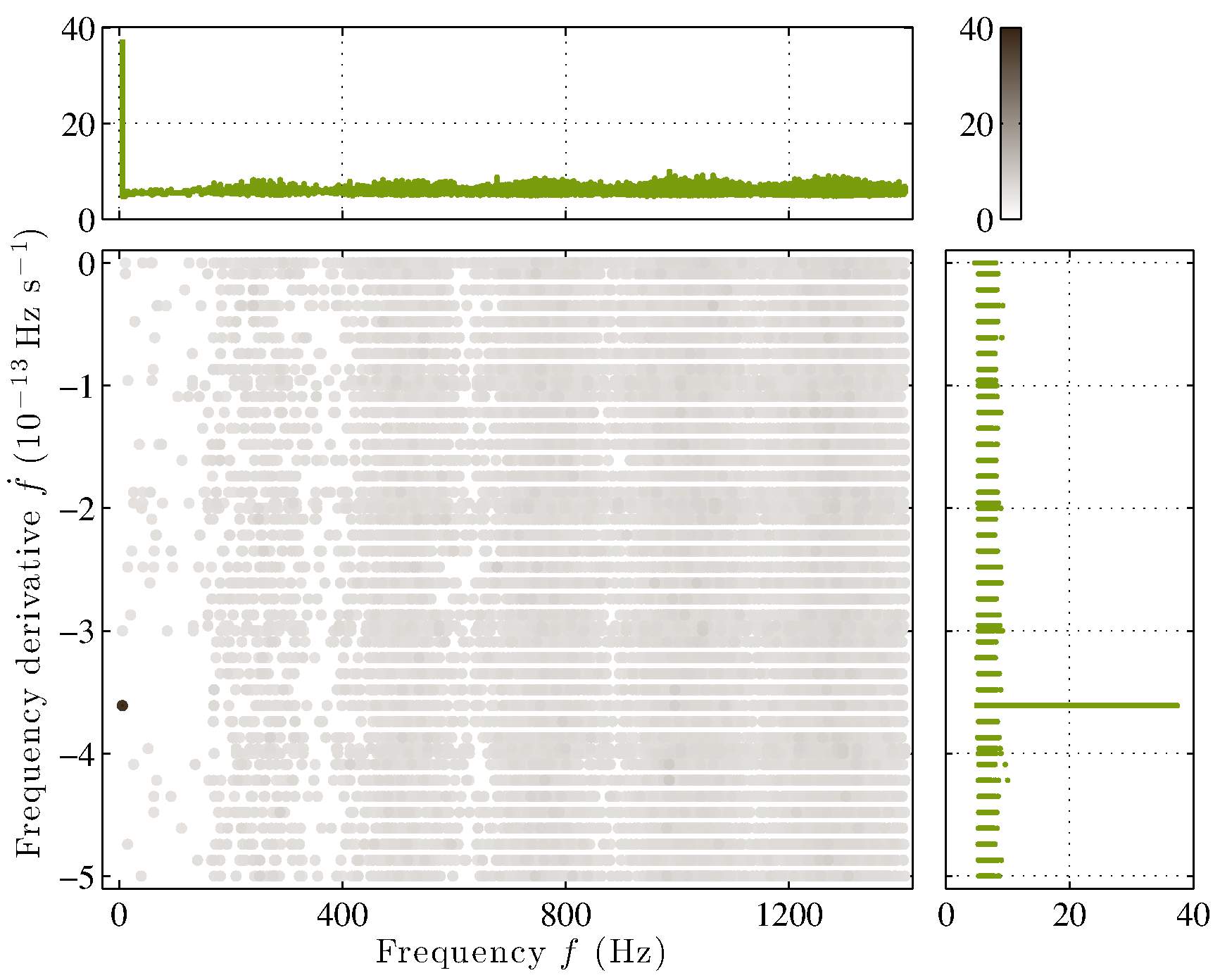}}\hspace{1em}
	\subfigure[\label{f:fullycohJ1620}]
       	{\includegraphics[width=\columnwidth]{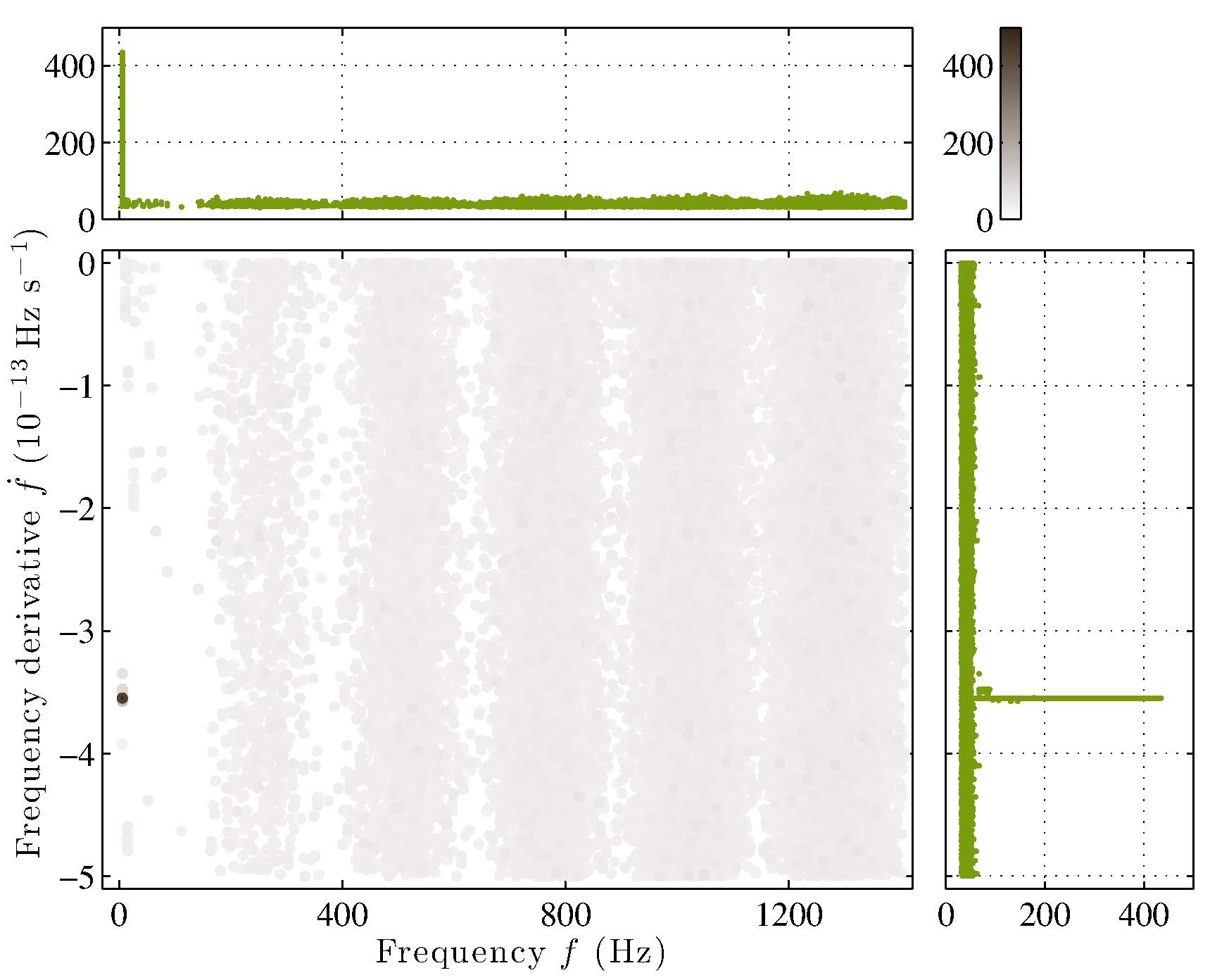}}
	\caption{Search results for the newly-discovered pulsar 
	\mbox{PSR~J1620--4927} (the large black dot in the lower-left of each panel).
	 The left panel \subref{f:semicohJ1620} shows the
          semi-coherent search results, representing about 2~CPU-years
          of computing on a single core. (Since the computing cost
          scales as the square of frequency, a search up to 64~Hz as in previous searches
          \citep{16gammapuls2009,8gammapuls2010}
          would have taken about 1.4~CPU-days on a single core.)  The
          bottom-left panel shows the semi-coherent detection
          statistic~$S$ as a function of $f$ and $\dot f$, maximized
          over the sky-grid.  The value of $S$ is represented by the colorbar.
          A further maximization over $f$ is shown to the
          right; a further maximization over $\dot f$ is shown above.
          \mbox{PSR~J1620--4927}, the darkest point near the bottom
          left of each figure, stands out clearly from the noise.
          In the same form, the right panel \subref{f:fullycohJ1620}
          presents the fully-coherent follow-up search results of the
          previous candidate (and every other candidate
          ``dot'') shown in the left panel.  The quantity plotted is
          now the fully-coherent detection statistic~$P$ over the
          entire data set. As explained in the text, for each
          candidate this covers a region of parameter space which is
          four steps of the semi-coherent grid in each dimension.}
\end{figure*}

To compute $S$ efficiently, a time series is constructed and
subsequently Fourier-transformed into the frequency domain.  The time
series contains $T\, \dfBW$ bins, where $\dfBW$ is the total frequency
bandwidth being searched at a time using complex
heterodyning\footnote{Complex heterodyning is a procedure which shifts
  frequencies in time-series data by a fixed offset $f_h$.  This is
  accomplished by multiplying the time series by $\rme^{-2\pi f_h
    t_j}$, shifting all frequencies by $f_h$.}  at the center of
$\dfBW$ \citep[see e.g.][]{Patel2010}.  The time series is initialized
to zero, then the values of \mbox{$w_j\, w_k\, \rme^{- i \pi \dot f \;
    (t^2_j-t^2_k) }$} are added into the bins determined by the time
differences $\Delta t_{jk} = t_j-t_k$, for all pairs of photons $j,k$
for which $0 < |\Delta t_{jk}| \le T$; the bin index is obtained by
rounding the absolute value of the product \mbox{$\dfBW\, \Delta
  t_{jk}$} to its nearest integer value.

Then the array is Fourier-transformed into the frequency domain 
(exploiting the FFT) to obtain $S$ over the entire $f$ grid.
Up to an overall window-dependent normalization, one can write for given values of
$f$, $\dot f$ and sky position,
\begin{equation}
  S = \sum_{j,k=1}^N Q(|\Delta t_{jk}|/T ) \; w_j\, w_k\, \rme^{-2\pi i
  f\Delta t_{jk}- i \pi \dot f \; (t^2_j-t^2_k) } ,
      \label{e:semidefinition3}
\end{equation}
where the rectangular function $Q(x)$ is unity if~\mbox{$0 < x \le 1$} and
vanishes otherwise\footnote{Note that by symmetry $S$ is real, because
interchanging indices $j$ and $k$ is equivalent to complex conjugating the 
exponential factor.}. 

Although other aspects are different, the use of an FFT applied to time differences 
is very similar to techniques previously used in blind searches of \textit{Fermi}-LAT
data \citep[ Equation~3 therein has a typo which is corrected in
  Equation~2 of \citet{Ziegler2008}]{TimeDiffTech2006}.  
The method used in \citet{TimeDiffTech2006} was the first application of this classic 
method \cite[e.g.][]{BlackmanTukey1958} to gamma-ray astronomy 
(in estimating the power spectrum an approximate autocovariance function is calculated
using a maximum lag, and then Fourier-transformed).

In contrast to previous searches, our method uses an optimal gridding of the 
parameter space for both the semi-coherent and coherent stages, as well as an 
automated follow-up, and incorporates the spin-down corrections in a way that 
permits heterodyning and highly-efficient code.

The search was done on the 1680-node \textit{Atlas} Computing Cluster \citep{AtlasArticleMPG}
built around four-core processors with 8~GB of random-access memory; 
for these we used a heterodyning bandwidth $\dfBW = 256$~Hz.  
Breaking the full frequency-range of the search into frequency-bands 
allows the computation to fit into memory, and also allows the use of 
different sky grids in each band.
This further reduced the computational cost, since the number of
required sky grid points increases with the square of frequency. 

After computing $S$ on the four-dimensional grid in parameter space,
points with statistically-significant values of $S$ are candidates for
possible pulsar signals, and are followed up in a second stage.  
This is done by ``refining the grid'' and increasing the coherent integration time.
This is an hierarchical scheme which is analogous to ``zooming'':
successively swapping microscope objectives for ones of higher
magnification, then re-centering the interesting point on the slide
\citep[see e.g.][]{cutler:2005,Houghpaper}.  In our case, this is done
by constructing the fully-coherent detection statistic~$P$ over the
\emph{entire} data set (or equivalently taking the Gaussian window-size $T \to \infty$ in \Eref{e:Pdefinition}) obtaining
 \begin{equation}
     P = \frac{1}{{\kappa}^2} \biggl| \sum_{j=1}^{N} w_j \; \rme^{-i \Phi(t_j)} \biggr|^2,
     \label{e:Pfullycoh}
\end{equation}
where for convenience we have normalized by the positive
constant~$\kappa$ given by
\begin{equation}
   {\kappa}^2 = \frac{1}{2} \sum_{j=1}^{N} w_j^2.
    \label{e:sigmadefinition}
\end{equation}
The computing cost to coherently follow up a single candidate is negligible in 
comparison to the cost of the previous semi-coherent search.

In selecting statistically-significant  semi-coherent candidates which 
are automatically followed up in the second stage using a fully-coherent analysis,
we do not use a fixed threshold to define ``statistical significance''. 
In the semi-coherent stage, the search code keeps an internal list of the strongest 
signal candidates.  Each member of this list is coherently followed up
and corresponds to the largest value of $S$ detected in eight
adjacent spin-down values for the entire heterodyning
frequency bandwidth  and a single sky point. 

The refined grid of the fully-coherent follow-up covers a
region of parameter space of size $(4 \Delta f) \times (4 \Delta {\dot
  f}) \times (4 \Delta \theta \times 4 \Delta \theta)$ when projected into the ecliptic plane.  
In other words, it covers a region whose volume is 256 times larger than the
volume of a fundamental cell in the original grid: its extent in each
dimension of parameter space is four grid-intervals.  The refined grid
has a spacing given by the previous formulae for $\Delta f$, $\Delta
{\dot f}$, and $\Delta \theta$, except that the coherence time~$T$ is
set equal to the length of the \emph{entire} data set, and $\gamma = 1$.
Since in this case only a small parameter-space region around the
candidate is explored, it is computationally-efficient to compute $P$
directly in the time-domain (FFTs are not used), exploiting the
sparsity of the photon data.

If the value of $P$, which measures the fully-coherent power (in a
single harmonic), is statistically significant, then in a third stage further refinement
is carried out using higher harmonics (Fourier-components). We adopt
the so-called $H$-test, which has been widely used in X-ray and
gamma-ray pulsar detection \citep{deJaeger1989,deJaeger2010}.  This test
measures the statistical significance of the energy in the first 20
(non-DC) Fourier-components of the pulse-profile as a function of
phase. As in \Eref{e:Pdefinition}, the $H$-test can also be 
modified to include the photon probability weights~$w_j$
\citep[see][]{KerrWeightedH2011}.  Note that Equation~(5) in
\cite{KerrWeightedH2011} contains an error; corrected formulae used
in this work are given below.

\begin{figure*}[t!!]
 \centering
 \subfigure[\label{f:HtestsigmaSky}]
  {\includegraphics[width=0.67\columnwidth]{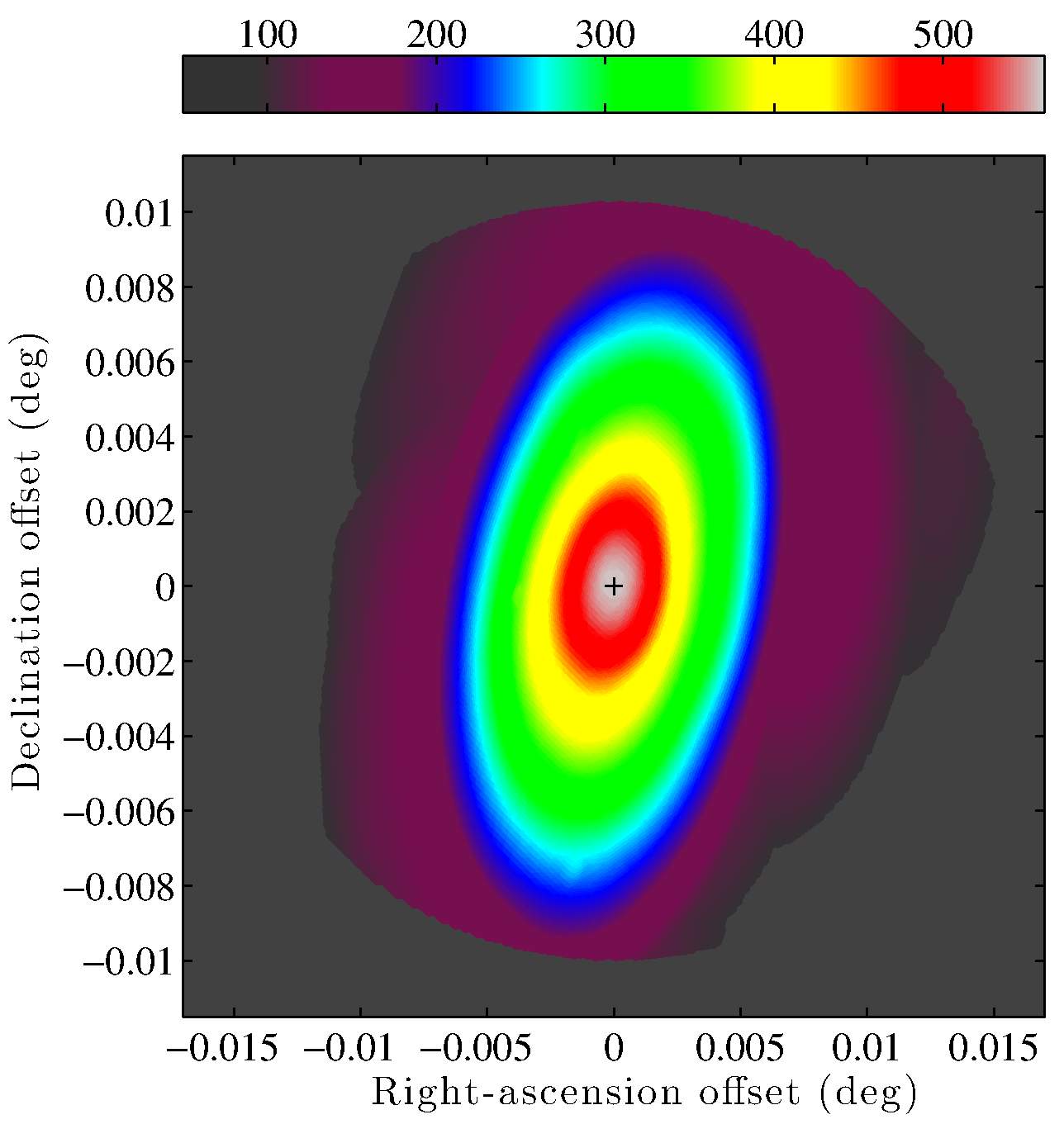}}\hspace{0.25em}
 \subfigure[\label{f:HtestsigmaFFdot}]
  {\includegraphics[width=0.67\columnwidth]{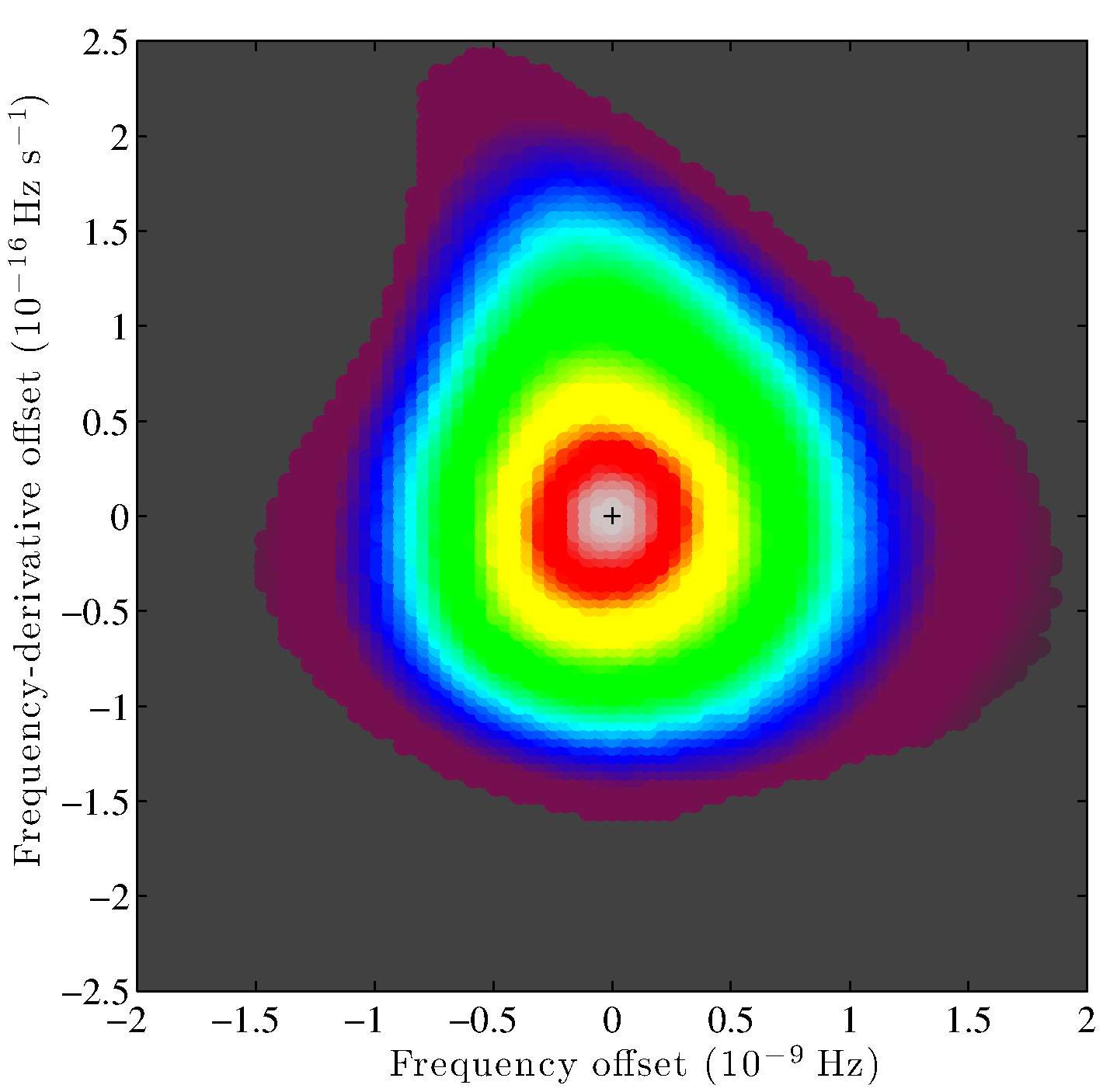}}\hspace{1em}
 \subfigure[\label{f:HtestTime}]
  {\includegraphics[width=0.63\columnwidth]{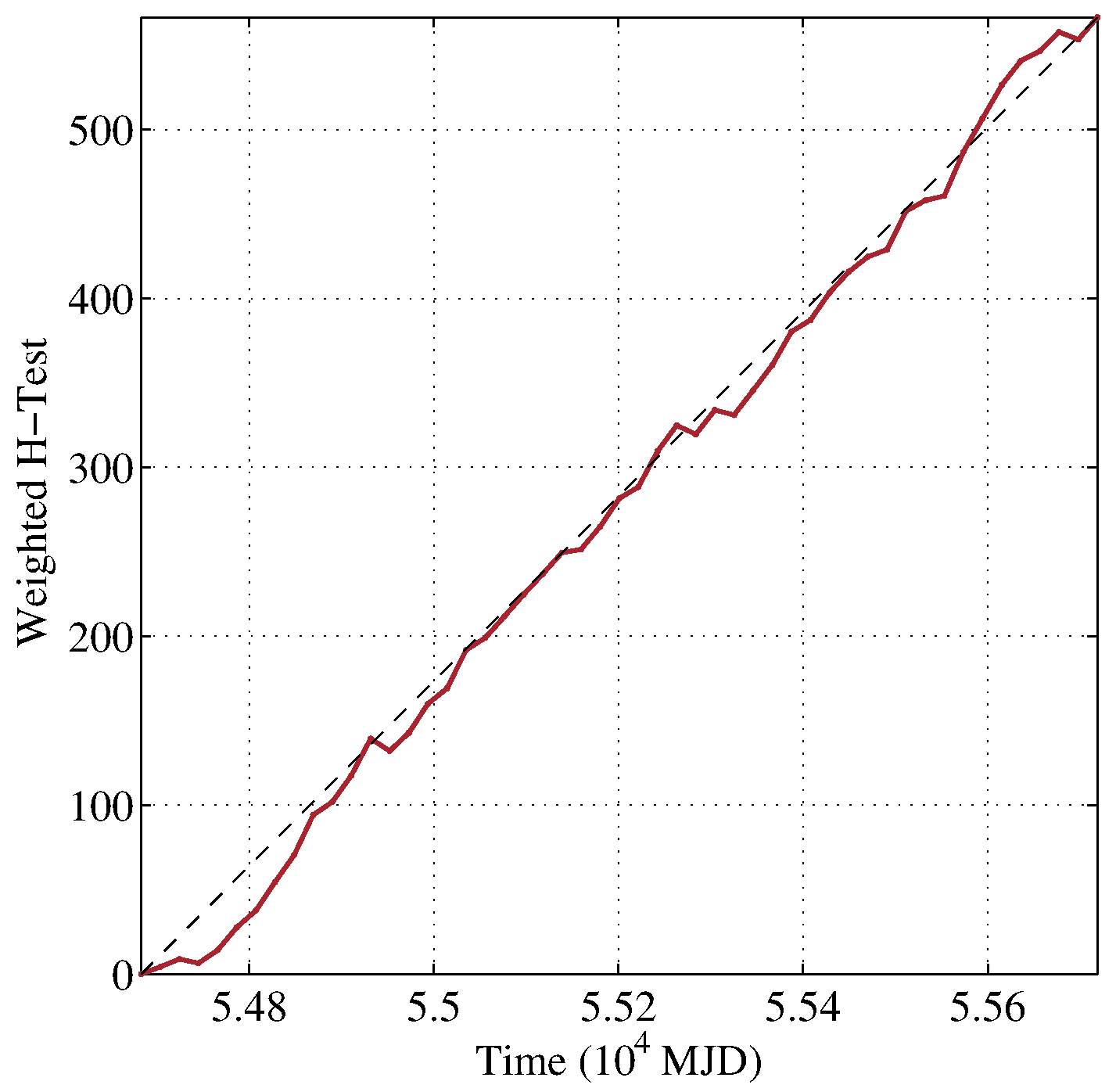}}
  \caption{\label{f:Htestsigma} 
    The weighted $H$-test statistic for \mbox{PSR~J1620--4927}.
    Panels~\subref{f:HtestsigmaSky} and \subref{f:HtestsigmaFFdot} show contour plots 
    of the weighted $H$-test statistic as a function of sky position~\subref{f:HtestsigmaSky} 
    and \mbox{$(f,\dot f)$} \subref{f:HtestsigmaFFdot}.
    These are peaked at the parameter-space location indicated by the black cross
    and the axes show the offset from these values.    
    Panel~\subref{f:HtestTime} shows how the maximum weighted $H$-test statistic    
    [shown in~\subref{f:HtestsigmaSky} and~\subref{f:HtestsigmaFFdot}] accumulates with time.    
    The $H$-test increases (approximately) linearly with time, 
    as expected for a pulsar that is emitting uniformly. 
    }
\end{figure*}

The weighted $H$-test statistic is defined as follows.  For each
photon arrival time $t_j$, the pulse phase~$x_j$ (between zero and
one) is calculated as \mbox{$x_j = \left[ \Phi(t_j) \mod 2 \pi \right] / 2\pi$}.
The pulse profile (for $0 \leq x < 1$) is a sum over the photons 
\begin{equation}
  p(x) = \sum_{j=1}^N w_j \, \delta(x-x_j),
\end{equation}
where $x_j$ is the pulse phase of the $j$th photon and $\delta(x)$ is a one-dimensional
Dirac delta-function. It can be expressed as a Fourier series
\begin{equation}
  p(x) = \kappa  \sum_{\ell= -\infty}^\infty \alpha_\ell \, \rm{e}^{2\pi i\ell x},
\end{equation}
which implies that the (complex) Fourier coefficients $\alpha_\ell$ are given by
\begin{equation}
  \alpha_\ell = \frac{1}{\kappa} \sum_{j=1}^N w_j \, \rm{e}^{-2\pi i \ell x_j},
\end{equation}
where the definition of $\kappa$ is identical to \Eref{e:sigmadefinition}.
The normalization of the Fourier coefficients has been chosen so that
{\it if} the photon arrival times are uniformly distributed,
independent random variables, {\it then} in the limit of large numbers
of photons, for $\ell>0$, $\Re(\alpha_\ell)$ and $\Im(\alpha_\ell)$
are independent Gaussian random variables with zero mean and unit
variance. Finally, the weighted \mbox{$H$-test} statistic is defined as
\begin{equation}
  H = \max_{1 \le L \le 20} \biggl[\sum_{\ell=1}^L \bigl|\alpha_{\ell} \bigr|^2 - 4(L -1)\biggr].
  \label{e:weightedHtest}
\end{equation}
The quantity which is subtracted, $4(L-1)$, is motivated by an
empirical numerical study \citep{deJaeger1989}, providing the best
omnibus test for unknown pulse profiles.

Maximizing $H$ over sky-position, frequency $f$, and
spin-down rate $\dot f$ selects the ``narrowest and sharpest''
overall pulse profile.  In contrast, maximizing the statistic \mbox{$P=|\alpha_1|^2$}
of \Eref{e:Pfullycoh} favors putting more power into the lowest harmonics.

Using $H$ as the test statistic, a further stage of parameter-space
refinement is done in the same way as before: in the frequency,
spin-down, and sky parameters, we cover regions that include four
grid-steps (in each dimension) of the previous grid.  The chosen
refinement factor (ratio of the number of grid points after and before
refinement, in each dimension) is about an order of magnitude.  
At each grid point, the weighted $H$-statistic of \Eref{e:weightedHtest} is
found. The parameter-space point with the largest statistic is
selected as our best estimate of the pulsars' parameters which
are then further refined through the timing-analysis procedure \citep{Ray2011}
described in Section~\ref{ssec:timing}.

The hierarchical search pipeline has been validated by successfully recovering 
previously known gamma-ray pulsars, including some of the
brightest gamma-ray MSPs \citep{FermiPSRCatalog,PletschGuillemot}.
In the next section, the complete search scheme is illustrated 
with a detailed example.

\section{Example: Results for PSR~J1620--4927}
\label{s:example}

Before giving the results for all of the pulsars that have been
discovered with this new search method, we first go through a single 
example in detail.  This illustration uses \mbox{PSR~J1620--4927},
the first new pulsar found in this work.

Figure~\ref{f:semicohJ1620} shows the first stage of the analysis.  For
each $f$ and $\dot f$ value, the largest value of $S$ found in the sky
grid is displayed as an intensity.  The point of highest intensity 
corresponds to PSR~J1620--4927's $f$ and $\dot f$ parameters.

The second analysis stage is the automated follow-up of all candidates 
shown in Figure~\ref{f:semicohJ1620}.  
As explained in Section~\ref{s:method}, this is accomplished by carrying 
out a fully-coherent search over a small region of parameter space around 
each candidate which has been identified as statistically significant in
the previous semi-coherent stage.  If the candidate found in the
semi-coherent step was simply a statistical outlier, then the
fully-coherent statistic $P$ will not be significant.
Figure~\ref{f:fullycohJ1620} presents the results of the
fully-coherent statistic $P$ for \mbox{PSR~J1620--4927}.  
Again the point of highest intensity compared to the background, 
indicating the presence of a coherent signal in the data set, 
is due to the new pulsar.

The importance of the photon probability weights $w_j$ can be
illustrated by repeating the analysis using the same 8000
photons with the weights set to unity.  (Note that the weights were
used in selecting the 8000 photons, so this is not a complete
comparison.) The result is that Figure~\ref{f:semicohJ1620}
still shows a statistically-significant outlier in the semi-coherent
search step, which is followed up automatically and gives a
statistically-significant outlier in the fully-coherent output of
Figure~\ref{f:fullycohJ1620}.  In both steps the signal and its
statistical significance are reduced, but the pulsar is still found.
However for some of the other new pulsars, this is not the case: 
if the weights are set to unity, then the pulsar is not detected.
Overall the weights play a larger role for sources 
in crowded regions of the sky, for example near the Galactic plane.

At the third stage, further refinement is carried out by maximizing
the weighted $H$-test statistic over a small region of parameter space
around the most significant candidate from the previous step.  As
described in Section~\ref{s:method}, the parameter-space grid used at
this stage is yet another order-of-magnitude finer than the one of the
previous stage. Figure~\ref{f:Htestsigma} shows the corresponding
results for \mbox{PSR~J1620--4927}.

The parameter-space point with the largest $H$-test statistic is
selected as our best estimate of the pulsar parameters at this stage. 
The uncertainties in these estimated parameters can
be obtained by using the fact that a 1$\sigma$ deviation for a Gaussian distribution 
has a value $\approx 0.6$ of the maximum.
In the current example, as shown in Figure~\ref{f:Htestsigma}, the value 
of the maximum weighted $H$-test statistic is~566.

Figure~\ref{f:HtestTime} illustrates the way that the weighted $H$-test
statistic accumulates over the total observation time.  The linearity
of this plot indicates that the properties of the source, and of the
data, were time-stationary; it shows that the pulsar's emission
is not changing with time.  In this way it also gives additional confidence 
that the pulsar is real and not a statistical noise outlier.  
(We have not provided analogous plots for the other eight pulsars 
reported in this paper, but they are similar.)

As shown in \citet{deJaeger2010} and \citet{KerrWeightedH2011}, the
weighted $H$-test has a false alarm probability approximately
described by \mbox{$P_{\textrm{FA}} \approx \rme^{-0.4\,H}$}.  For
this example, the associated false alarm probability is
\mbox{$\log_{10} P_{\textrm{FA}} \approx -98$}. We have not tried to
rigorously estimate the trials factor, but since the total number of
floating-point computations that could be performed by our computing
systems over a period of some months is less than $10^{21}$, we
conclude that \mbox{PSR~J1620--4927} is a real gamma-ray pulsar and
not a statistical outlier.

\section{Results for all new pulsars }
\label{s:allothers}

Table~\ref{t:pulnames} shows the names and sky positions of the nine 
discovered pulsars, and also lists known source associations.
The inferred rotational parameters of the new pulsars are presented in
Table~\ref{t:pulspinparams}. In addition, further derived parameters are given, 
including the spin-down power~\mbox{$\dot E = -4\pi^2 I f \dot{f}$}, where $I=10^{45}$~g~cm$^2$ 
is the assumed neutron star moment of inertia, 
and the magnetic field strength at the neutron star surface 
\mbox{$B_{\textrm{S}} \approx 3.2\times10^{19} (-\dot{f}/f^3 \,\textrm{s}^{-1})^{1/2}$~G} 
and at the light cylinder 
\mbox{$B_{\textrm{LC}}\approx 2.94\times10^{8} (-\dot{f} f^3\,\textrm{s}^5)^{1/2}$~G}, 
respectively.

\begin{figure}
	\centering
	\includegraphics[width=\columnwidth]{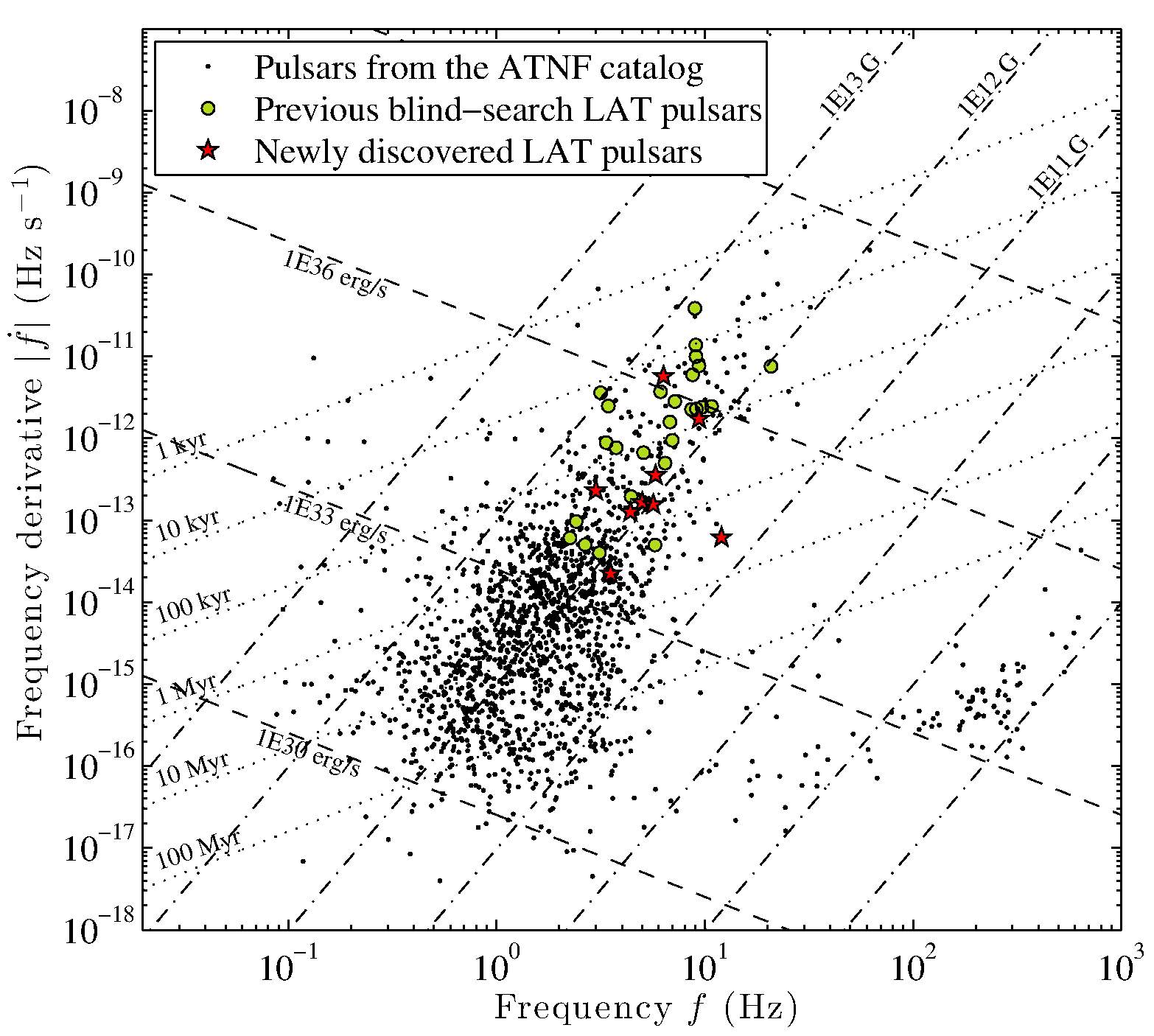}
	\caption{\label{f:ffdotplot} The frequency $f$ and
          spin-down rate~$\dot f$ of the pulsar population.  The black dots
          show the approximately 1800~pulsars in the ATNF catalog
          \citep{ATNFcat}, excluding pulsars in globular clusters. 
          The green dots show the 26 gamma-ray pulsars discovered 
          in previous blind \textit{Fermi}-LAT searches
          \citep{16gammapuls2009,8gammapuls2010,2gammapuls2011}.  
          The nine red stars show the newly-discovered gamma-ray pulsars
          reported in this paper.  The dotted lines indicate constant
          characteristic ages~$\tau$, the dashed lines show contours
          of constant spin-down power~$\dot E$ and the dashed-dotted
          lines signify contours of constant surface magnetic field
          strength~$B_{S}$.}
\end{figure}

Figure~\ref{f:ffdotplot} plots the newly discovered pulsars on a \mbox{$f$--$\dot f$} diagram, 
where they can be compared with the known pulsar population.  One can
see that the new pulsars lie in a similar region (and hence belong at
large to the same population) as the previously-found blind-search
\textit{Fermi}-LAT gamma-ray pulsars.

Two of the nine new systems, \mbox{PSRs~J1803--2149} and J2111+4606, 
are young energetic pulsars (\mbox{$\dot E \gtrsim 6\times 10^{35}$ erg s$^{-1}$} 
and \mbox{$\tau \lesssim 100$ kyr}), located near the
Galactic plane.  Among the seven remaining less-energetic and older
pulsars, five are also located near the Galactic plane ($|b| < 5\degr$) 
and two are found at higher Galactic latitudes ($|b| > 10\degr$).  
One of these, \mbox{PSR~J0106+4855}, has the largest characteristic
age~$\tau = 3$~Myr and the lowest surface magnetic field
strength ($B_{\textrm{S}} \approx 2\times10^{11}$G) of all LAT blind-search
pulsars found to date. Also standing out, PSR~J2139+4716 has the smallest
spin-down power (\mbox{$\dot E = 3\times10^{33}$~erg s$^{-1}$})
among all non-recycled gamma-ray pulsars ever detected \citep[cf.][]{DeLuca2011}.

\subsection{Source associations}

As listed in Table~\ref{t:pulnames}, all but one of the new pulsars
have gamma-ray counterparts from the \emph{Fermi}-LAT First Source
Catalog~\citep[1FGL,][]{FermiFirstSourceCatalog}.  The exception is \mbox{PSR
  J2028+3332}, which first appeared in the 2FGL catalog.  In addition, 
these eight 1FGL unassociated sources have been classified as likely 
pulsar candidates based strictly on a comparison of their gamma-ray 
properties with previously detected LAT pulsars \citep{1FGLUnassoc}.

PSR~J2111+4606 is associated with the source  \mbox{0FGL J2110.8+4606} 
from the \emph{Fermi} Bright Source List \citep{FermiBSL}.
Only ten objects from this catalog still remain to be identified. 

The new system PSR~J1620--4927 is associated with the unidentified \emph{AGILE} 
source \mbox{1AGL J1624--4946} \citep{AGILEcat}. This \emph{AGILE} source is however 
about ten times brighter, with a flux of \mbox{(67 $\pm$ 13) $\times$ 10$^{-8}$ ph cm$^{-2}$ s$^{-1}$}. 
In addition, it has an error radius of 0$\fdg$58 and encloses other 2FGL sources. 
The new gamma-ray pulsar could therefore only explain a fraction of the total flux detected 
by \emph{AGILE}  from \mbox{1AGL J1624--4946}. 

While not formally associated in the 2FGL catalog \citep{FermiSecondSourceCatalog}, 
PSR~J2028+3332 lies within the 99\% error contours for the EGRET source \mbox{3EG J2027+3429}. 
However, the pulsar accounts for only about a fifth of the flux for the EGRET source. 
Another nearby LAT source, \mbox{2FGL J2025.1+3341}, is associated with an AGN 
and has a peak LAT flux that can account 
for the remainder of the measured EGRET flux.

We have searched the TeVCat online catalog of TeV sources\footnote{\url{http://tevcat.uchicago.edu/}}   
for very high-energy counterparts to the newly-discovered pulsars, 
but found no associations.  This is not surprising, considering that most of the new gamma-ray 
pulsars have moderate spin-down power (\mbox{$\sim$ 10$^{34}$ erg s$^{-1}$}) and 
relatively large characteristic ages ($\gtrsim$ 100~kyr), while pulsars associated with 
TeV Pulsar Wind Nebulae (PWNe) tend to be young and energetic. 
\mbox{PSRs J1803--2149} and \mbox{J2111+4606} have properties that make associations 
with TeV PWNe plausible; however, no detections of TeV emission from these directions 
have been reported.

\subsection{Timing analysis\label{ssec:timing}}

For each new gamma-ray pulsar the definitive parameters as given in
Tables~\ref{t:pulnames} and~\ref{t:pulspinparams} are determined
via precise measurements of pulse Times Of Arrival (TOAs) and
by fitting the parameters of a timing model to these measurements
using methods described in \cite{Ray2011}.

To obtain a set of TOAs, photons are extracted for each source using a
radius and minimum energy cut to optimize the signal-to-noise ratio
for each pulsar. Then these data are subdivided into segments of about
equal length. The set of best-guess pulsar parameters from the search
(maximizing~$H$) is used as a preliminary ephemeris to fold the photon
arrival times, producing a set of pulse profiles. The TOAs are then
measured by cross-correlating each pulse profile with a multi-Gaussian 
or kernel-density template derived from fitting the entire data set. 
This is done using the unbinned maximum likelihood method described 
in \citet{Ray2011}.  Then \textsc{Tempo2}~\citep{Tempo2} is used to fit 
the TOAs to a timing model including sky position, frequency and 
frequency derivative.

As often is the case, the youngest object found in this work 
(PSR~J2111+4606, $\tau=17$~kyr), exhibits particularly large timing noise. 
In turn, this requires including higher-order frequency-derivative 
terms to make a good timing model fit; in this case up to the fourth derivative of 
the frequency as shown in Table~\ref{t:pulspinparams}.
For the same reason, the timing models in Table~\ref{t:pulspinparams} 
for PSRs~J2030+4415 and J1803--2149  include terms up to 
the third and second frequency derivative, respectively.

\begin{figure}
\centering
	\hspace{-0.2cm}
		\subfigure
		{\includegraphics[width=0.49\columnwidth]{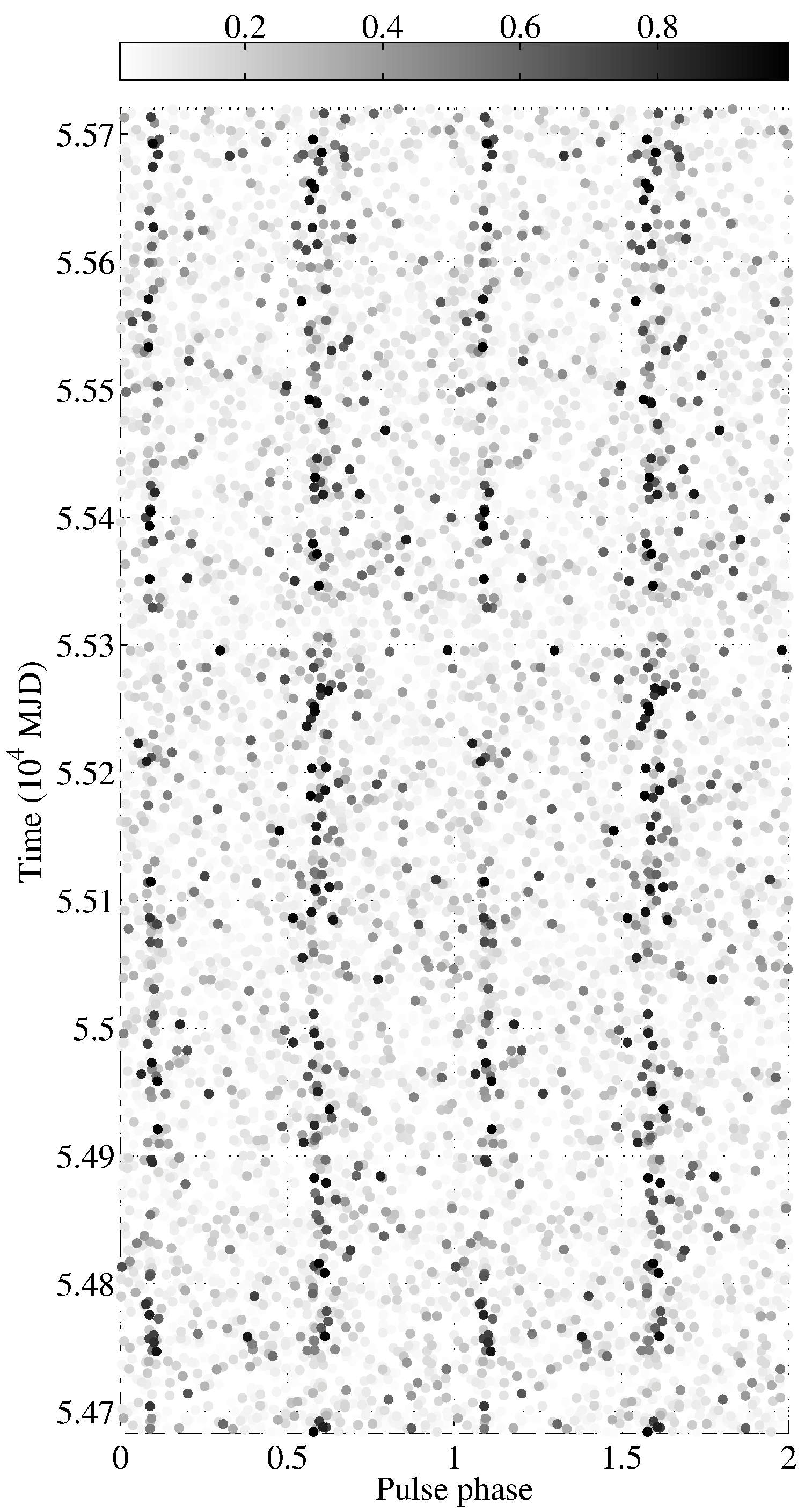}}
		\subfigure
		{\includegraphics[width=0.49\columnwidth]{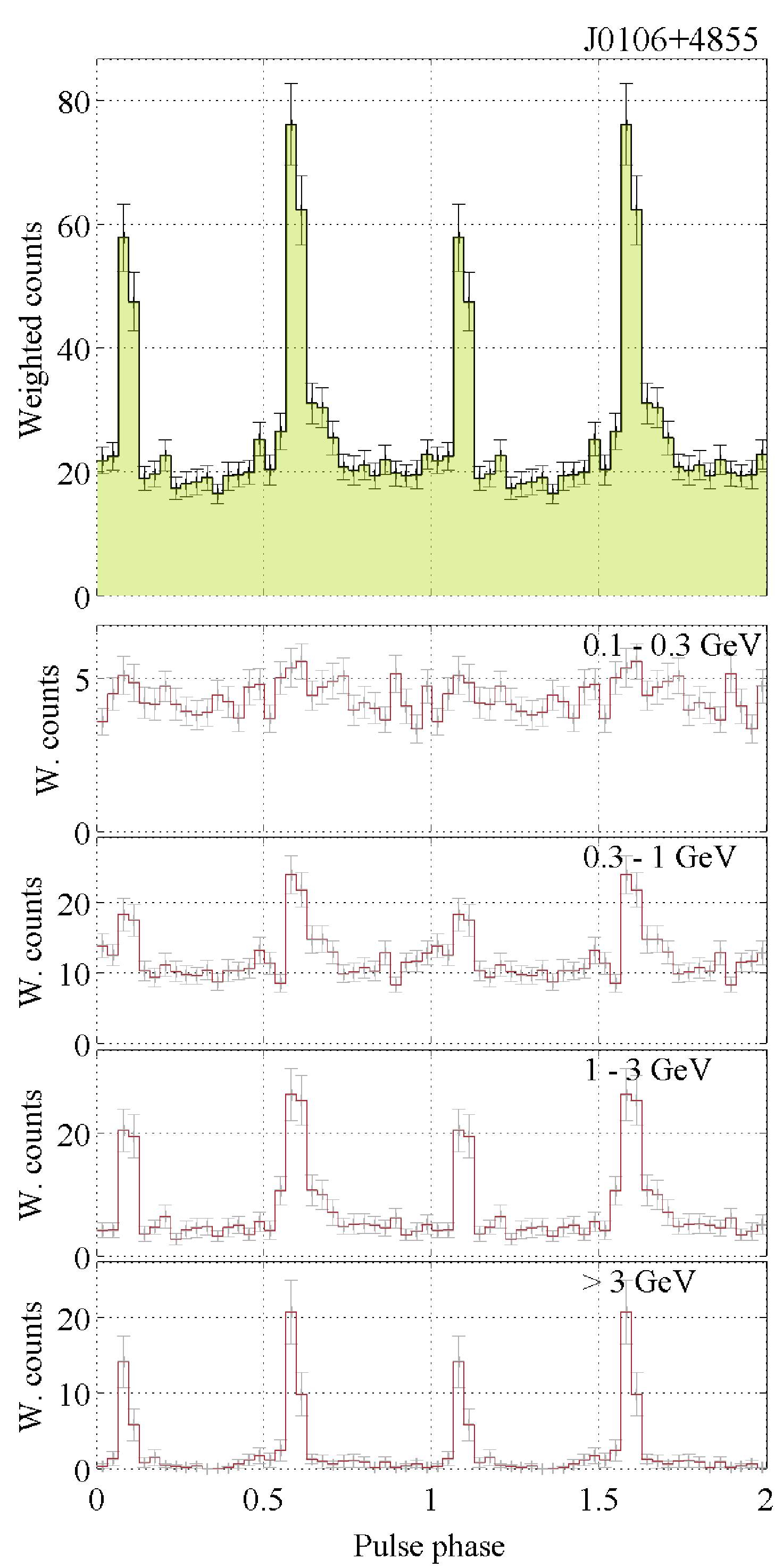}}
	\caption{\label{f:newpulsar-J0106} 
	 Phase-time diagram and pulse profile for \mbox{PSR J0106+4855}. 
	 The left panel shows the pulse phase at the arrival time of each photon, 
	 where the gray-scale intensity represents the photon probability weight. 
          The upper right plot shows the summed probability weights: 
          the integrated pulse profile using a resolution of 32 bins per rotation. 
          The error bars represent the 1$\sigma$ statistical uncertainties.
          The four plots below resolve the integrated pulse profile according to separate energy ranges.
          For clarity, the horizontal axis shows two pulsar rotations in each diagram.
          In obtaining these plots the 8000 events with the highest probability weights
          have been used (as in the blind search). }
\end{figure}

\begin{figure}
\centering
	\hspace{-0.2cm}
		\subfigure
		{\includegraphics[width=0.49\columnwidth]{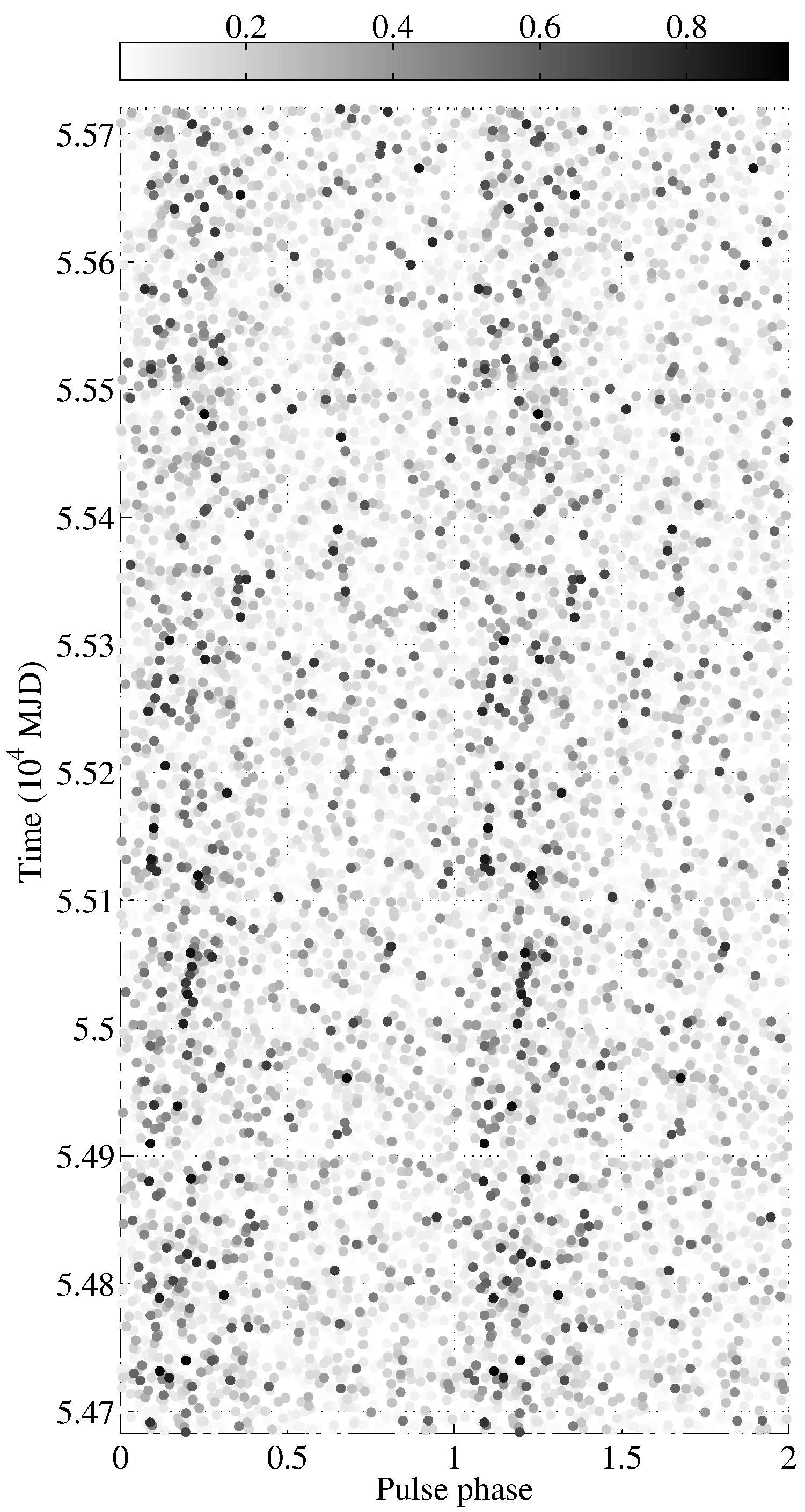}}\hspace{0.1cm}
		\subfigure
		{\includegraphics[width=0.49\columnwidth]{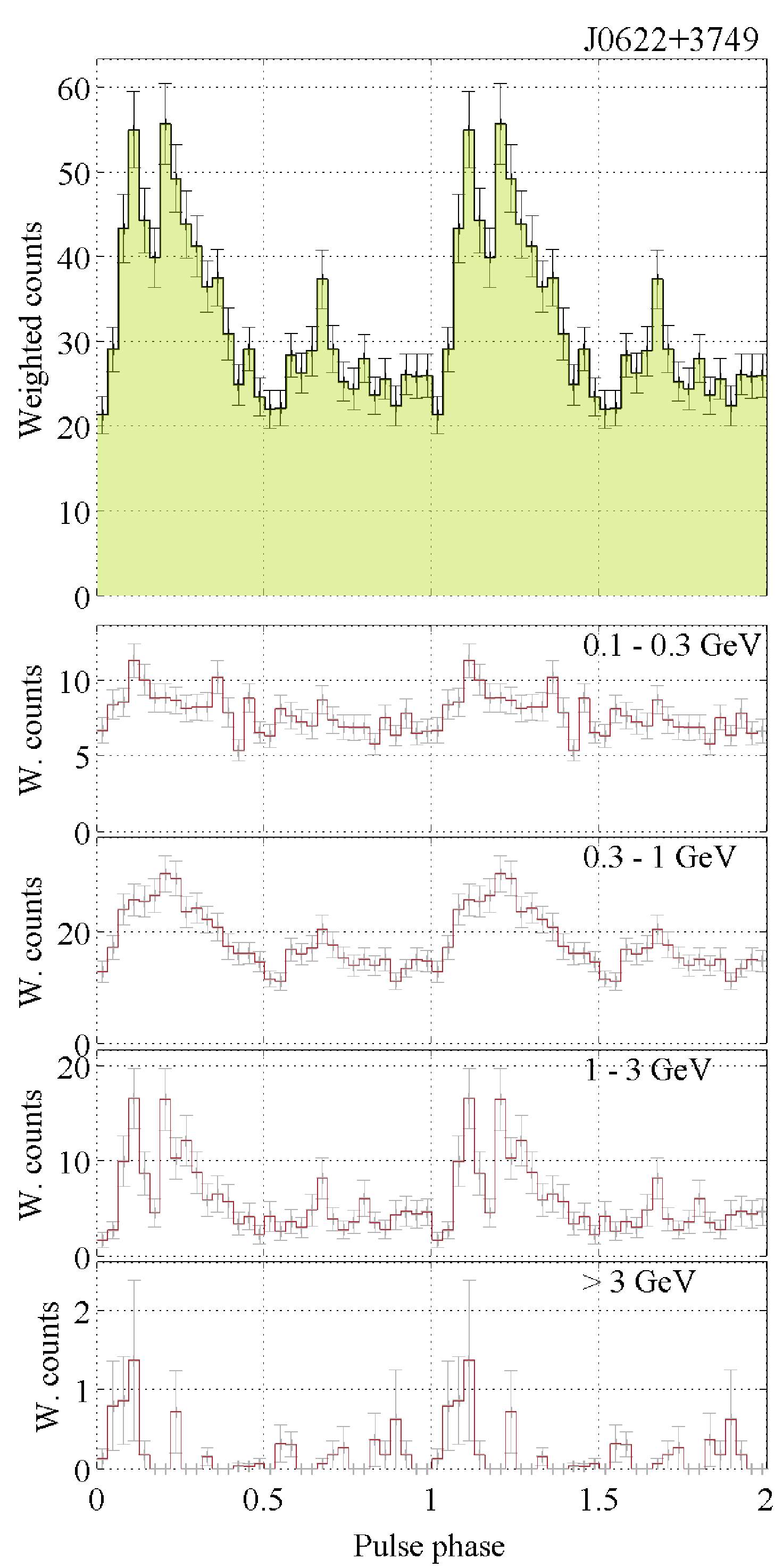}}
	\caption{\label{f:newpulsar-J0622} 
	 Phase-time diagram and pulse profile for \mbox{PSR J0622+3749}. 
	 The plots have identical form as those shown in Figure~\ref{f:newpulsar-J0106}.}
\end{figure}

\begin{figure}
\centering
	\hspace{-0.2cm}
		\subfigure
		{\includegraphics[width=0.49\columnwidth]{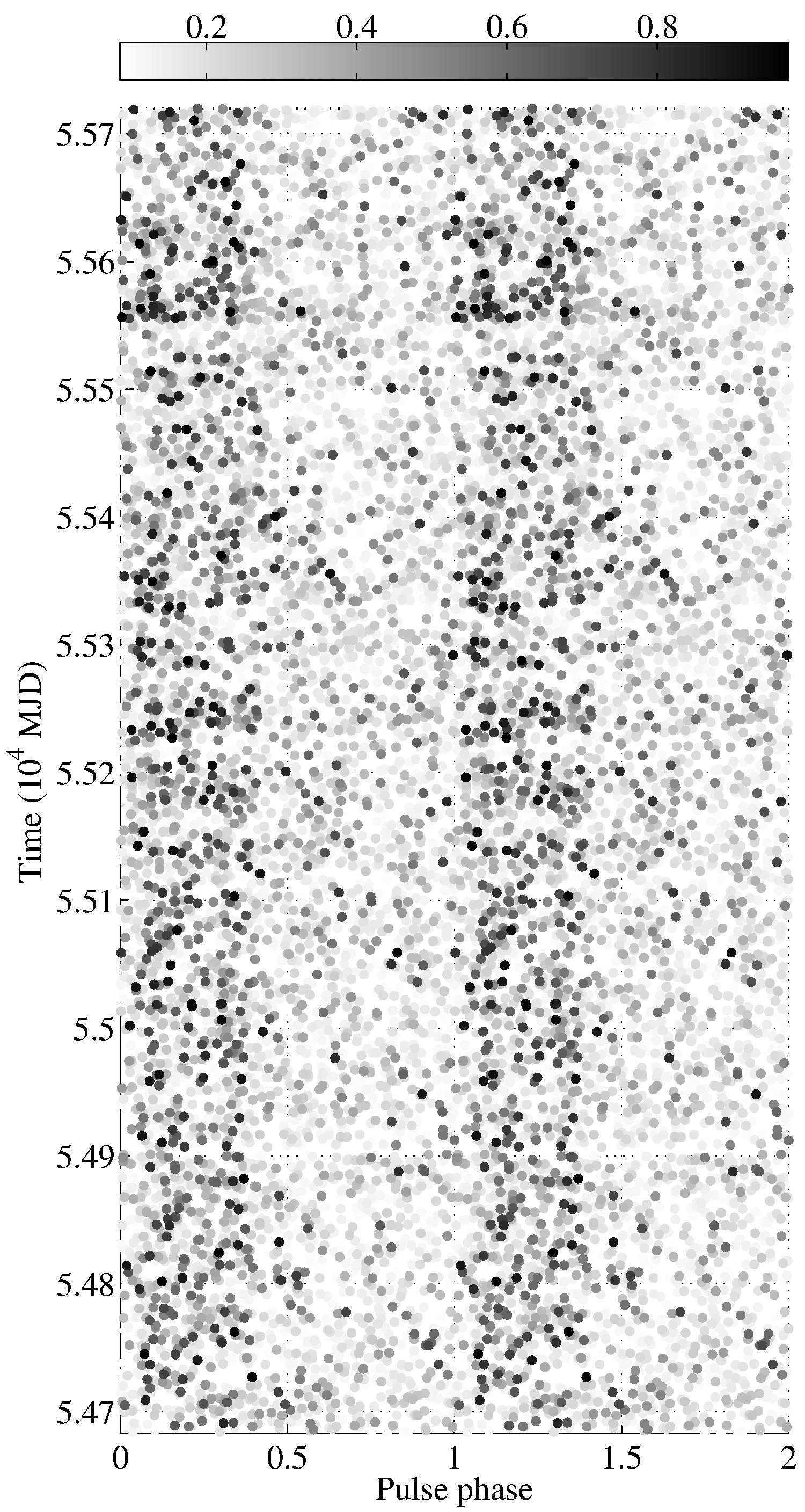}}\hspace{0.1cm}
		\subfigure
		{\includegraphics[width=0.49\columnwidth]{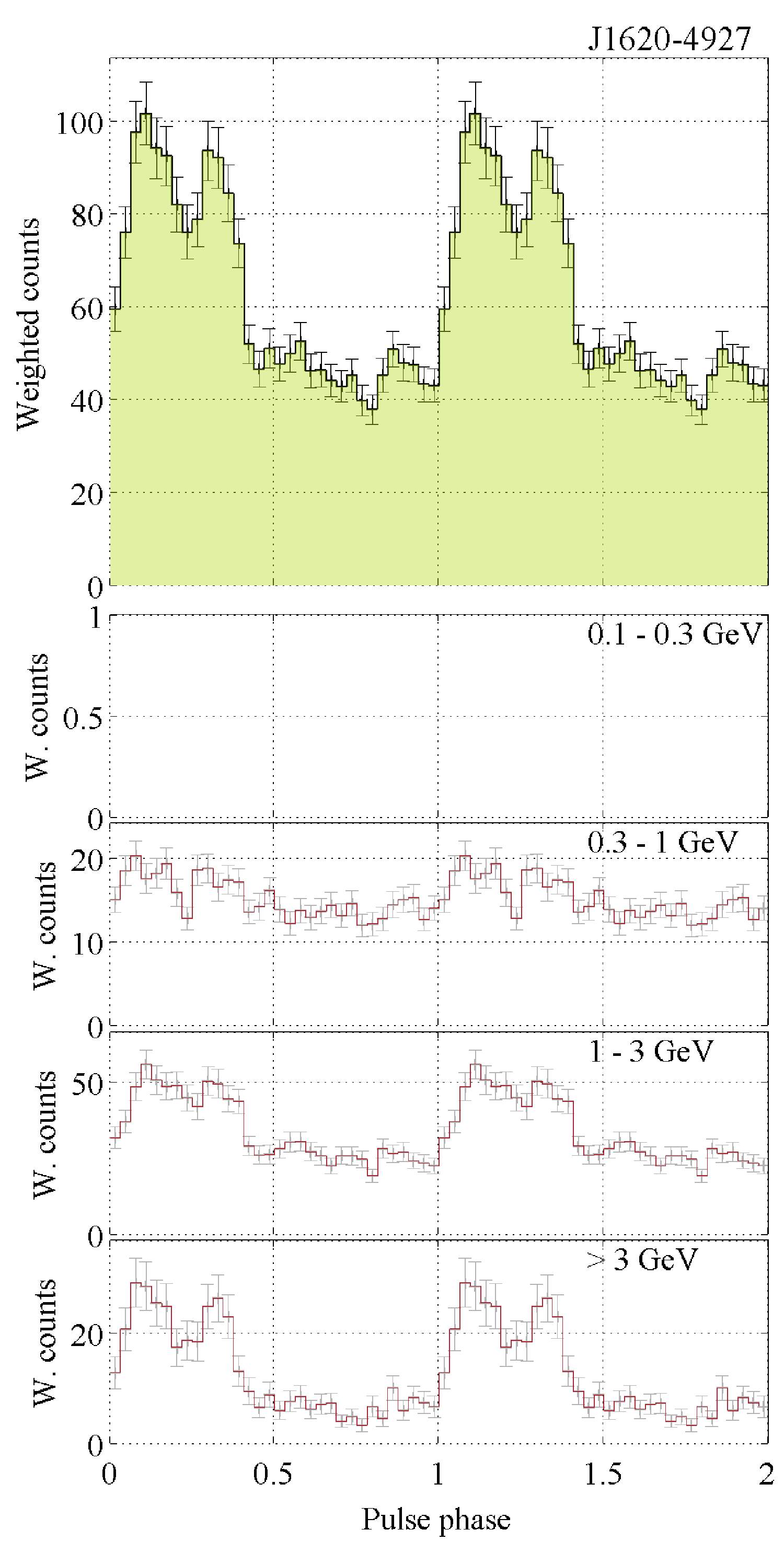}}
	\caption{\label{f:newpulsar-J1620} 
	 Phase-time diagram and pulse profile for \mbox{PSR J1620--4927}.
	 The plots have identical form as those shown in Figure~\ref{f:newpulsar-J0106}.
	 Note that for this particular pulsar, based on the photon probability weights, 
	 no selected events (among the 8000) have energies in the \mbox{0.1--0.3 GeV} range.}
\end{figure}

\begin{figure}
\centering
	\hspace{-0.2cm}
		\subfigure
		{\includegraphics[width=0.49\columnwidth]{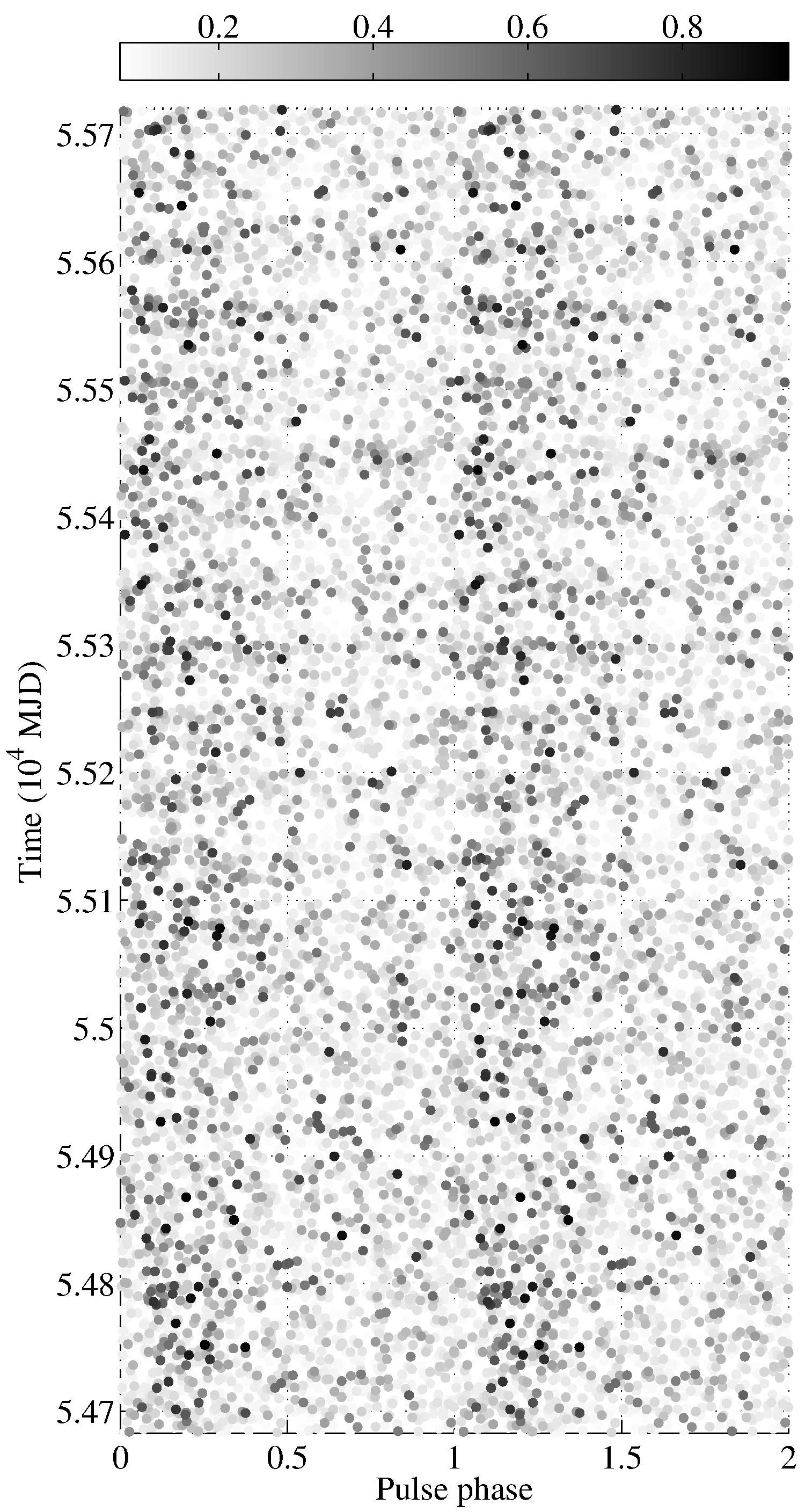}}\hspace{0.1cm}
		\subfigure
		{\includegraphics[width=0.49\columnwidth]{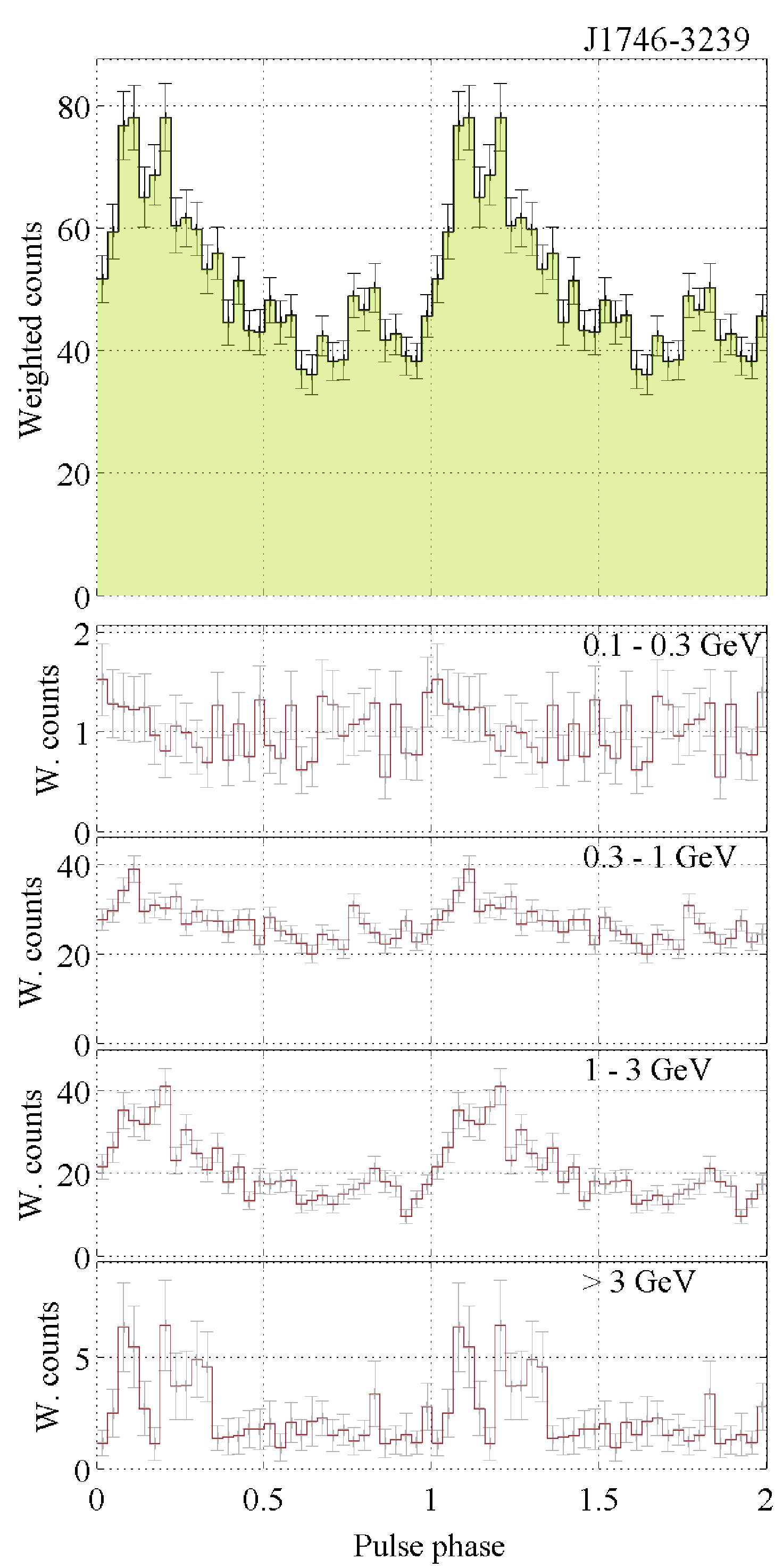}}
	\caption{\label{f:newpulsar-J1746} 
	 Phase-time diagram and pulse profile for \mbox{PSR J1746--3239}.
	 The plots have identical form as those shown in Figure~\ref{f:newpulsar-J0106}.}
\end{figure}

\begin{figure}
\centering
	\hspace{-0.2cm}
		\subfigure
		{\includegraphics[width=0.49\columnwidth]{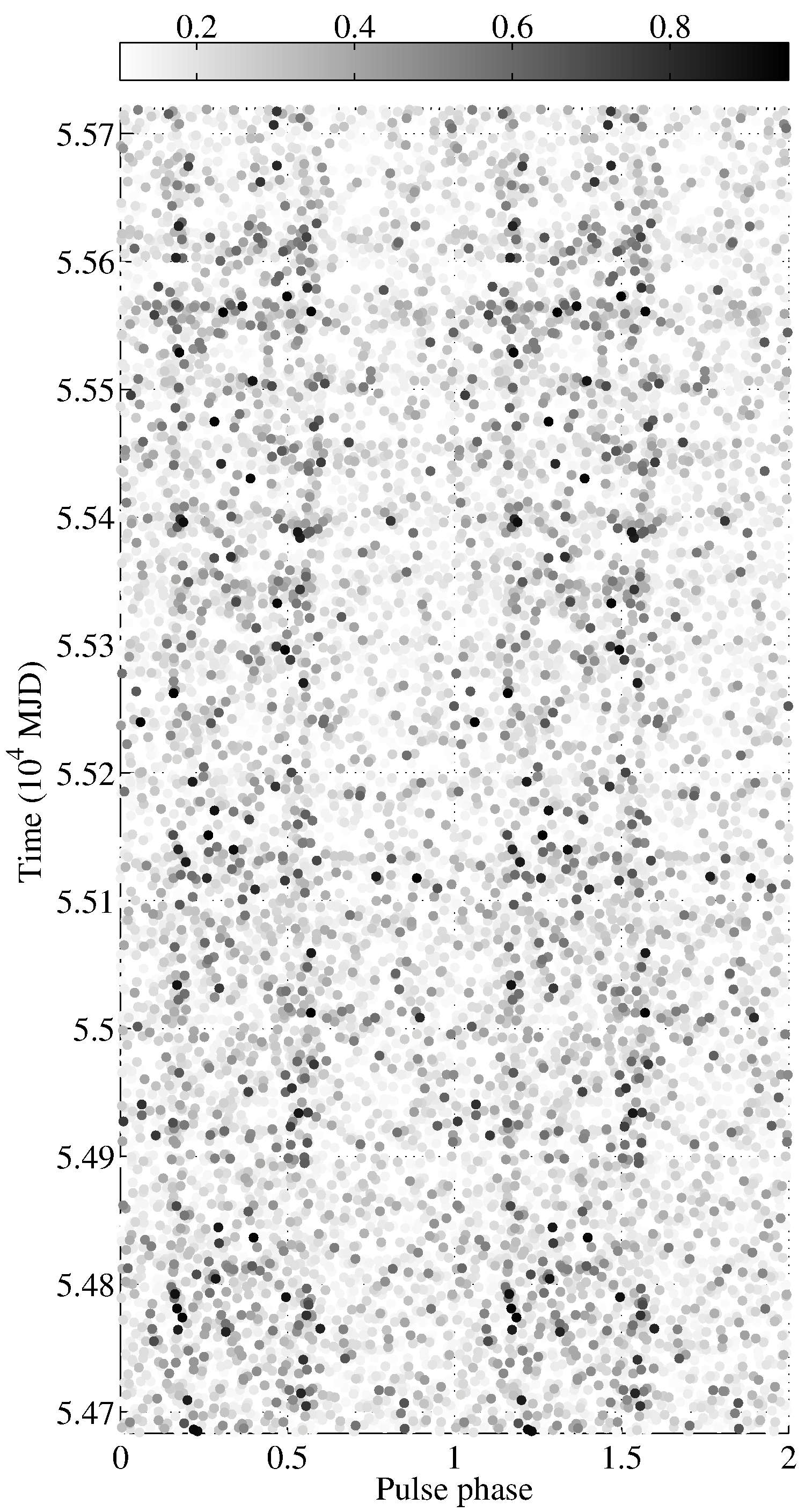}}\hspace{0.1cm}
		\subfigure
		{\includegraphics[width=0.49\columnwidth]{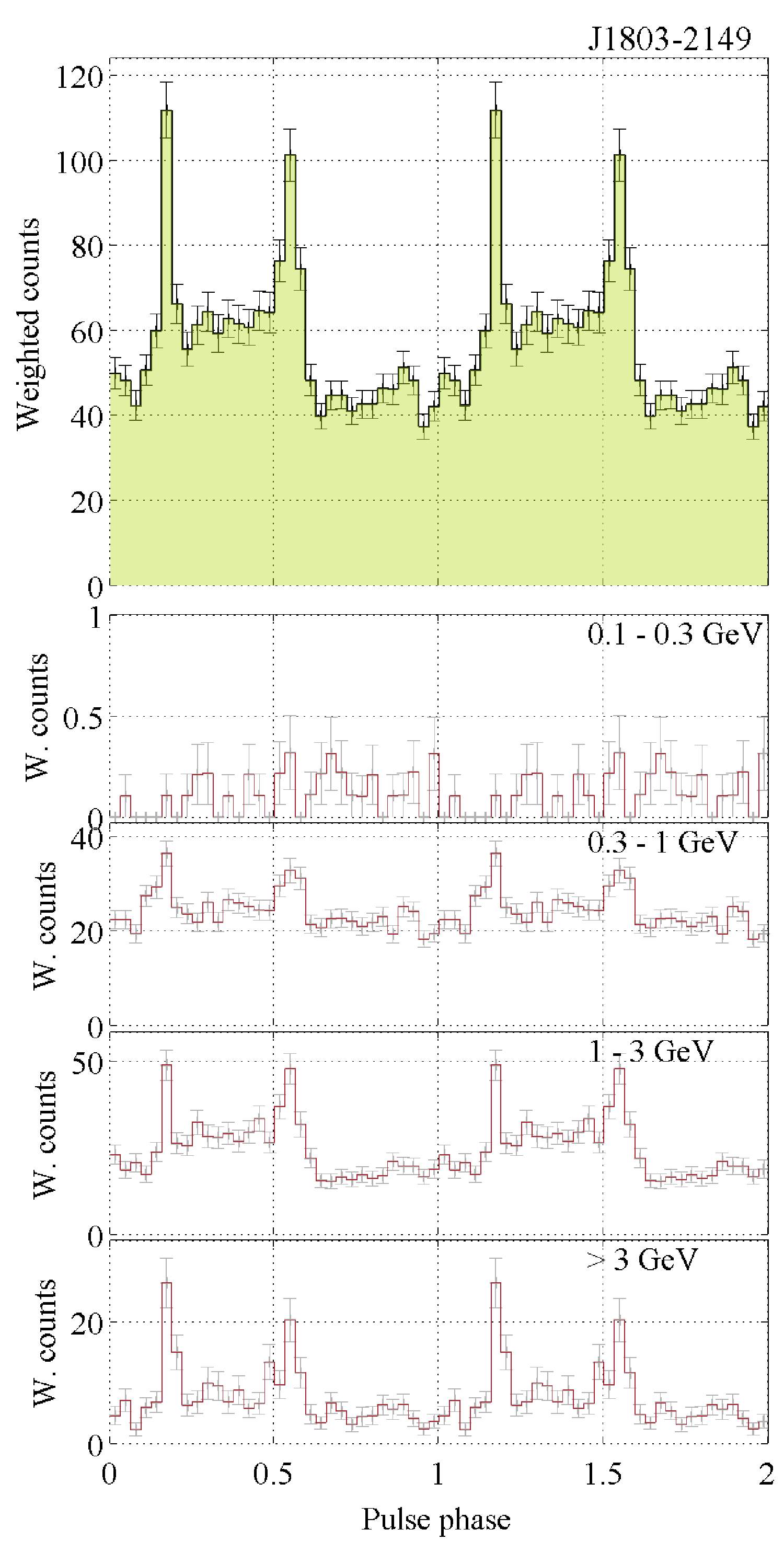}}
	\caption{\label{f:newpulsar-J1803} 
	Phase-time diagram and pulse profile for \mbox{PSR J1803--2149}.
	The plots have identical form as those shown in Figure~\ref{f:newpulsar-J0106}.}
\end{figure}

\begin{figure}
\centering
	\hspace{-0.2cm}
		\subfigure
		{\includegraphics[width=0.49\columnwidth]{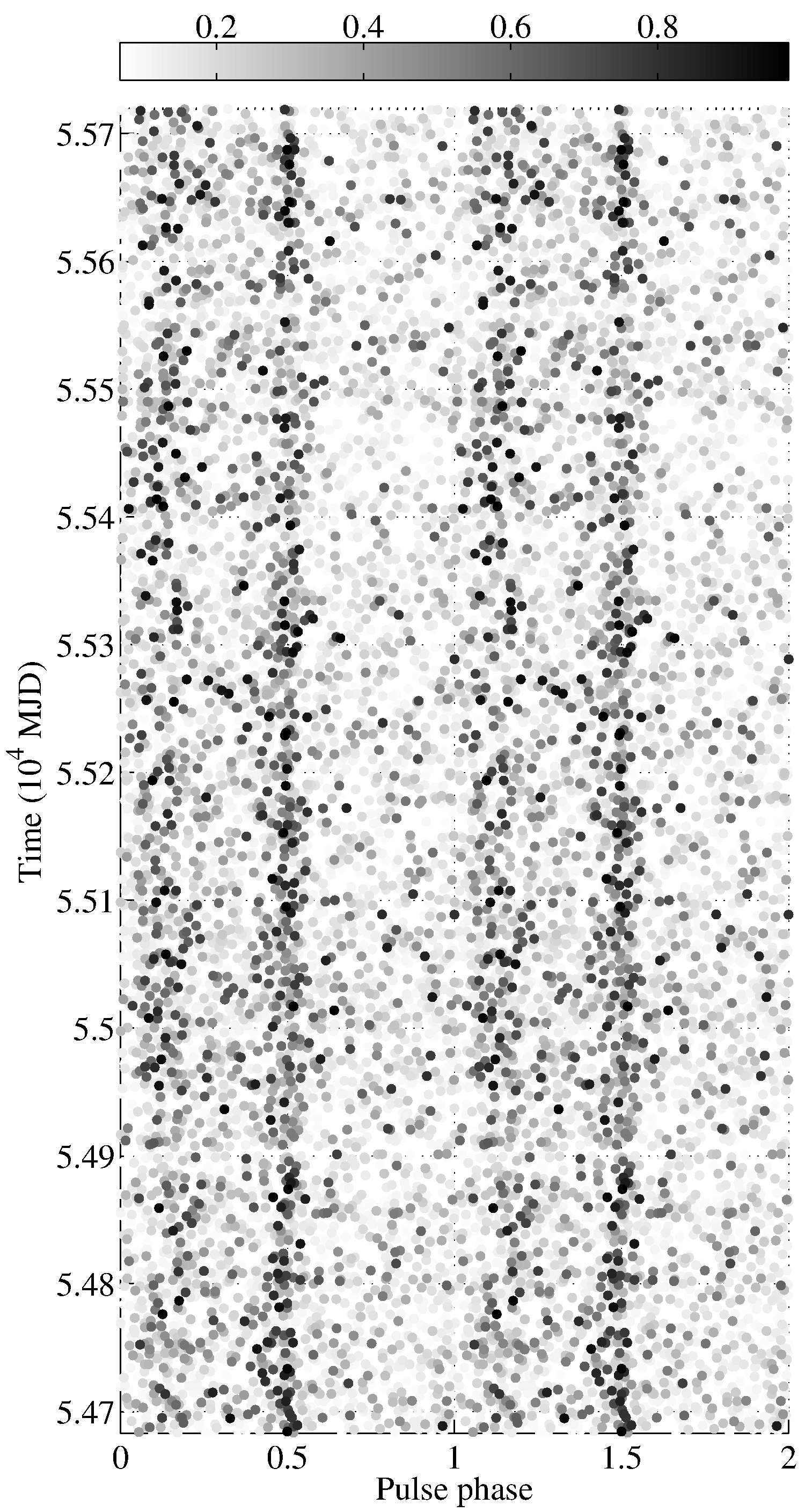}}\hspace{0.1cm}
		\subfigure
		{\includegraphics[width=0.49\columnwidth]{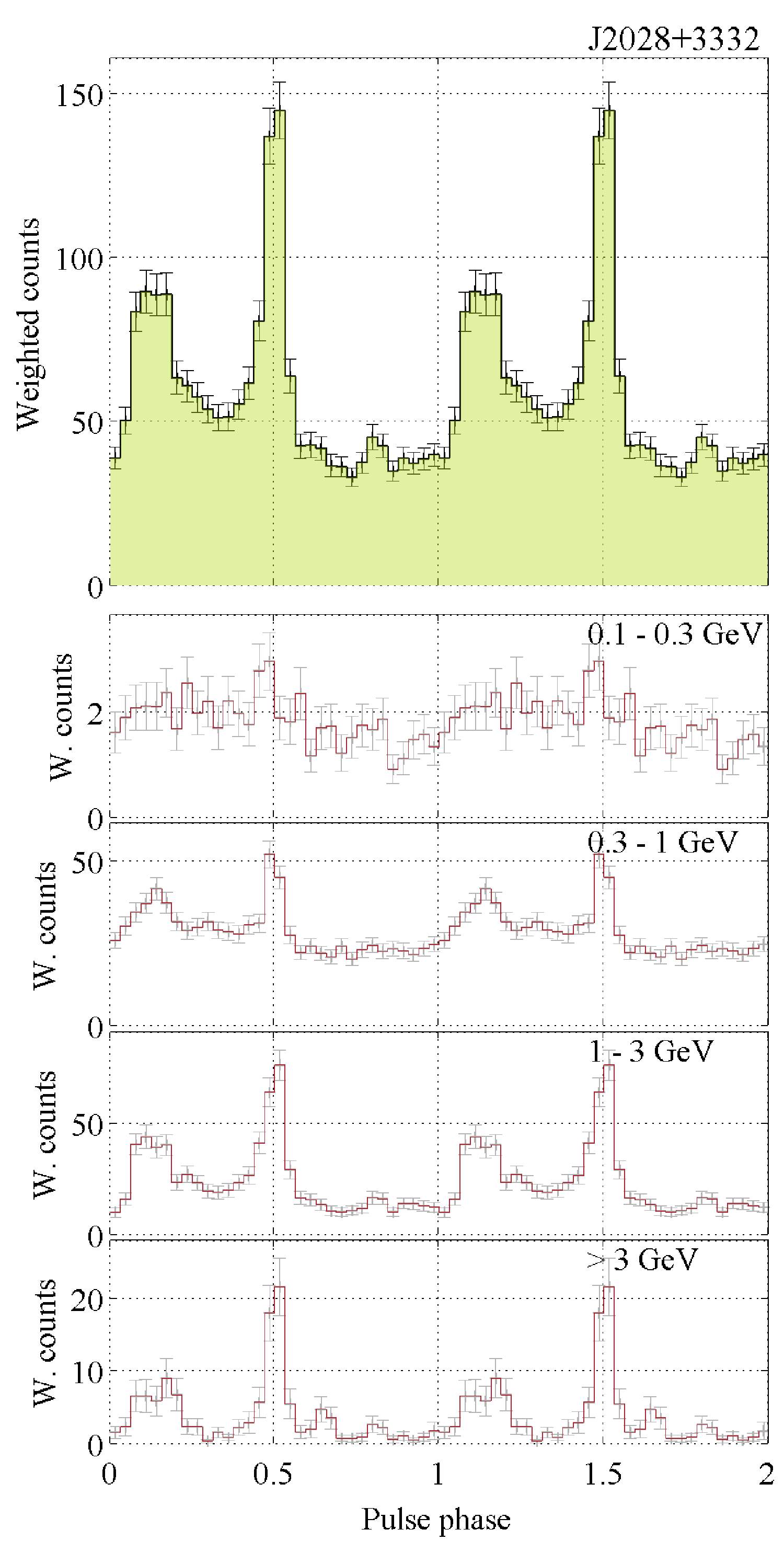}}
	\caption{\label{f:newpulsar-J2028} 
	 Phase-time diagram and pulse profile for \mbox{PSR J2028+3332}.
	 The plots have identical form as those shown in Figure~\ref{f:newpulsar-J0106}.}
\end{figure}

\begin{figure}
\centering
	\hspace{-0.2cm}
		\subfigure
		{\includegraphics[width=0.49\columnwidth]{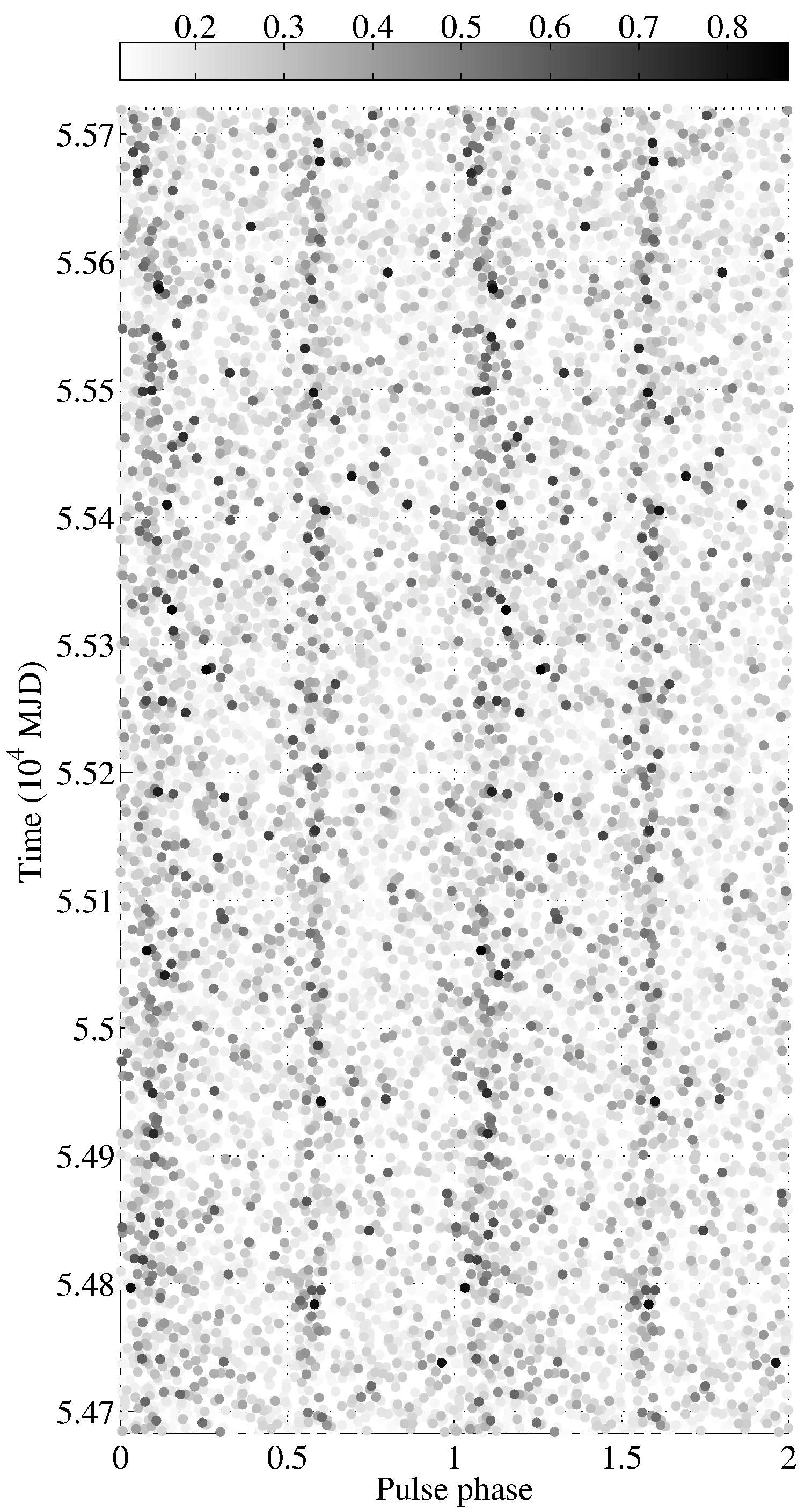}}\hspace{0.1cm}
		\subfigure
		{\includegraphics[width=0.49\columnwidth]{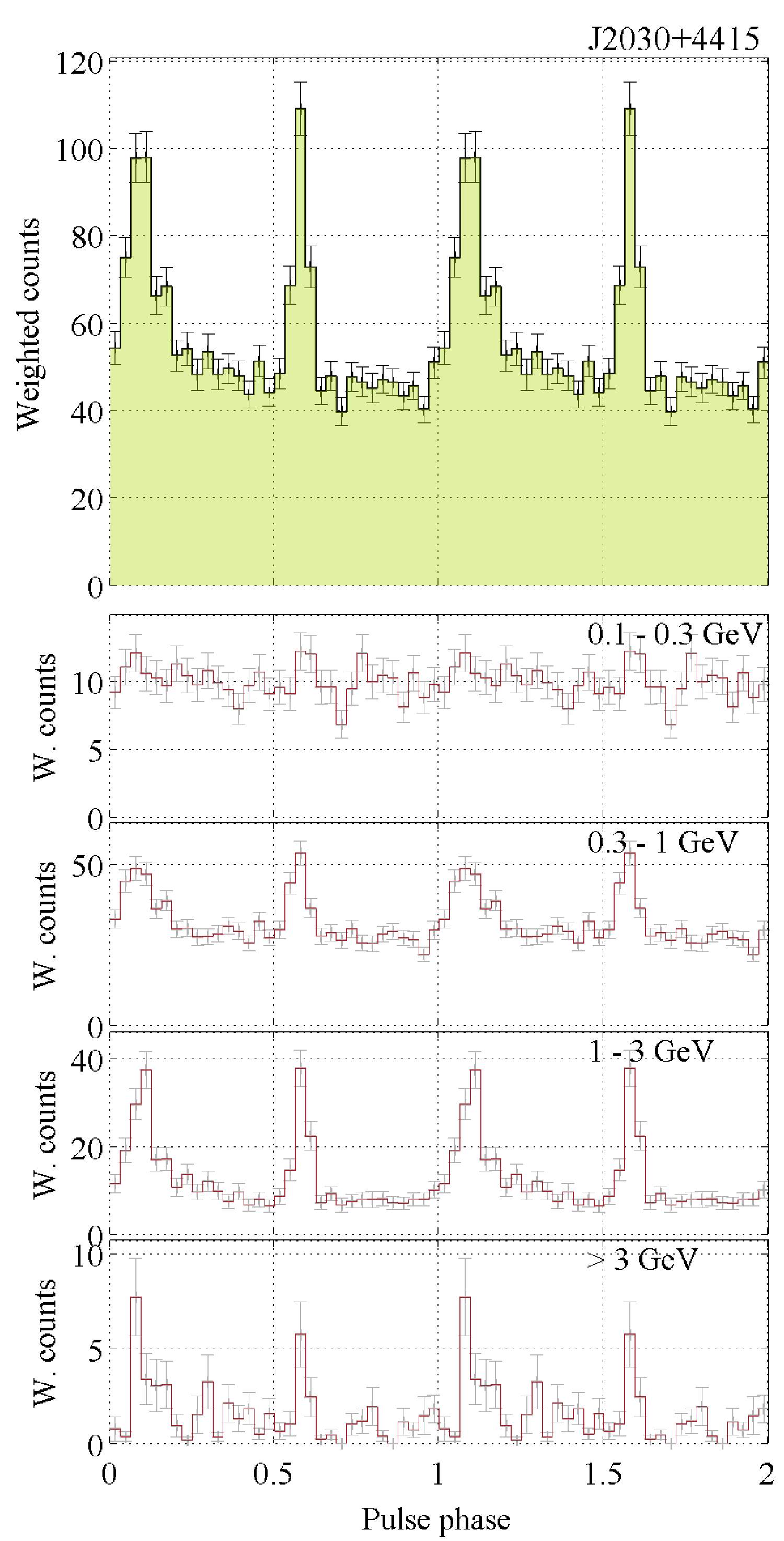}}
	\caption{\label{f:newpulsar-J2030} 
	 Phase-time diagram and pulse profile for \mbox{PSR J2030+4415}.
	 The plots have identical form as those shown in Figure~\ref{f:newpulsar-J0106}.}
\end{figure}

\begin{figure}
\centering
	\hspace{-0.2cm}
		\subfigure
		{\includegraphics[width=0.49\columnwidth]{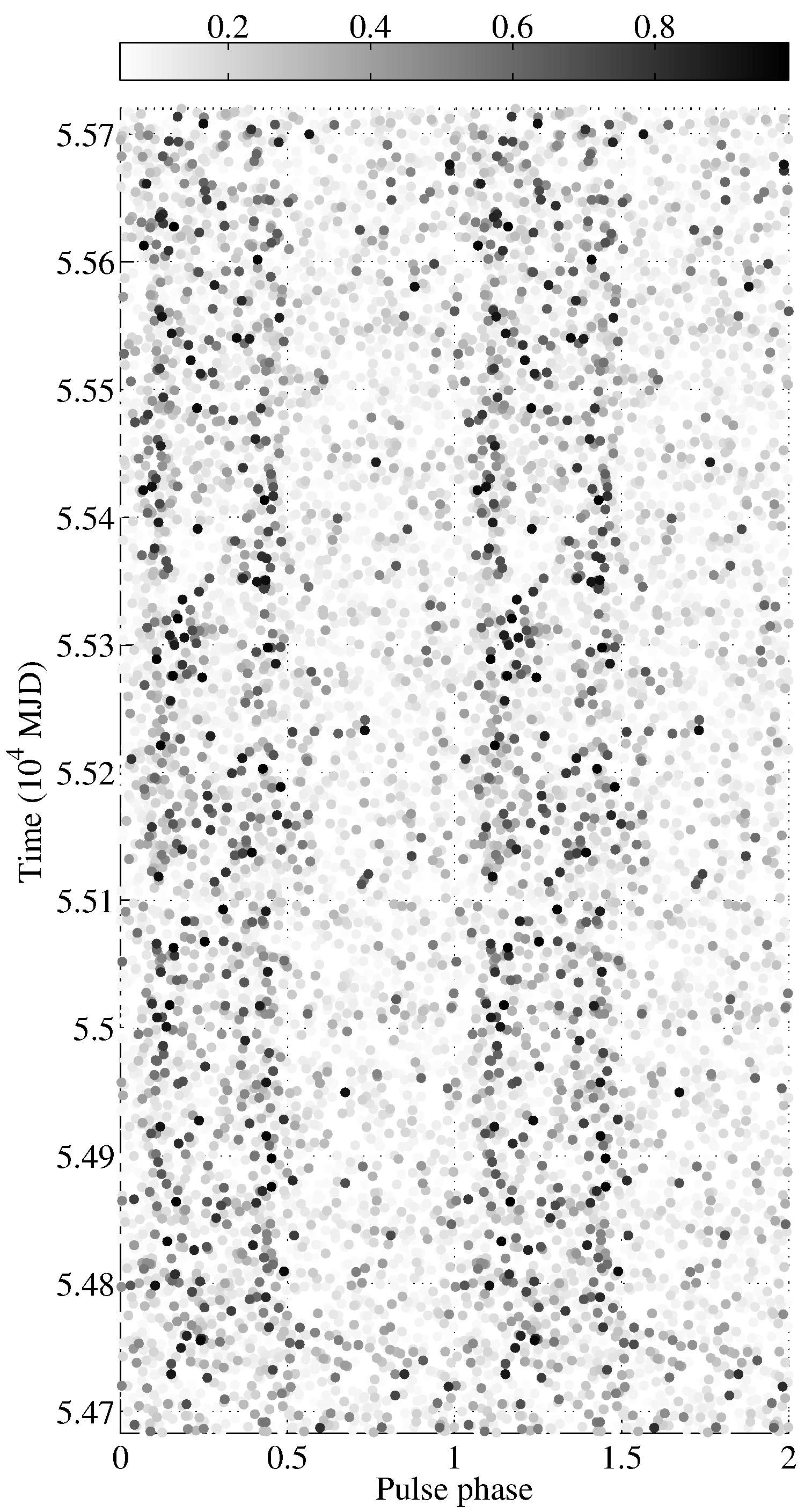}}\hspace{0.1cm}
		\subfigure
		{\includegraphics[width=0.49\columnwidth]{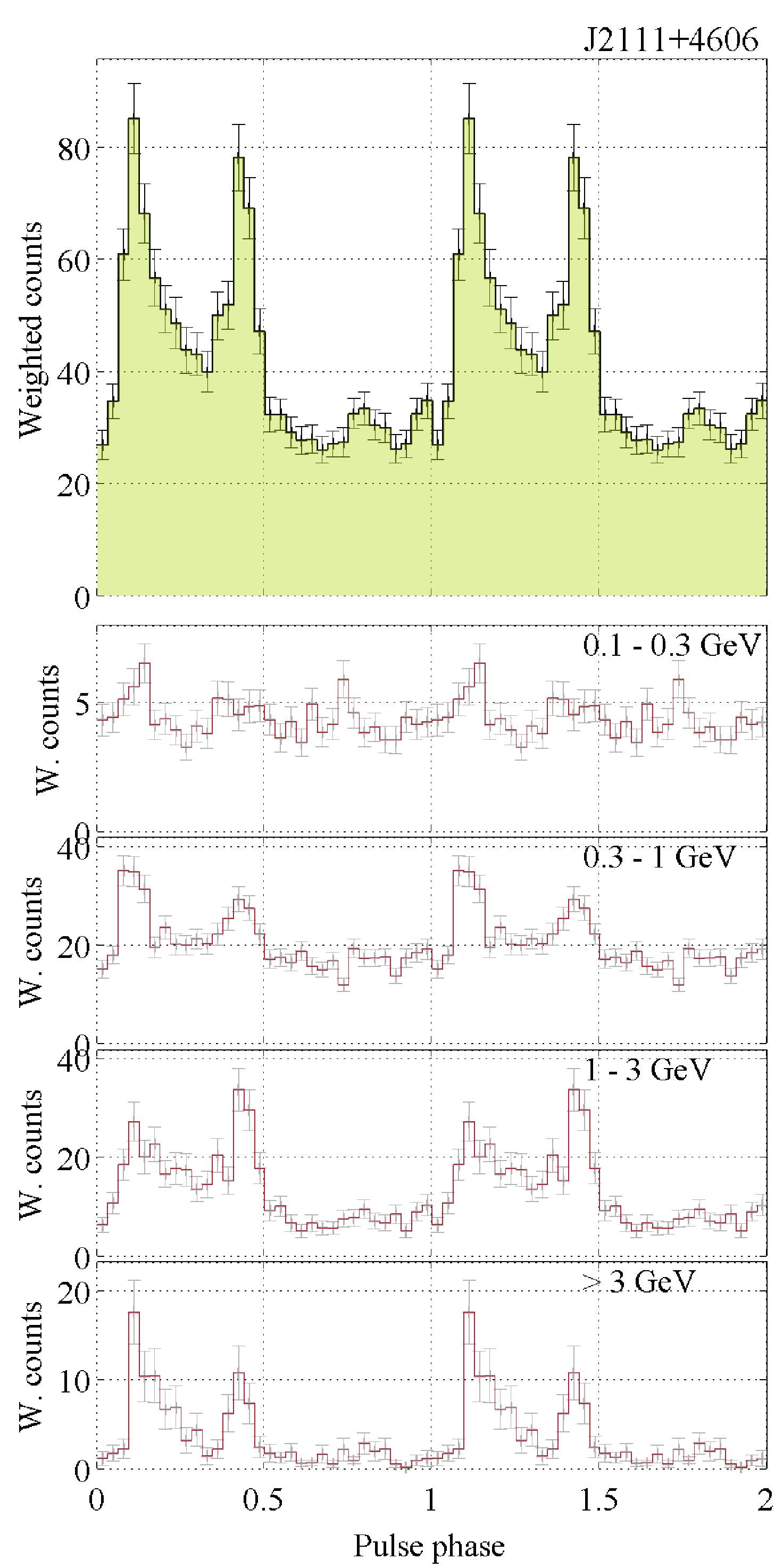}}
	\caption{\label{f:newpulsar-J2111} 
	 Phase-time diagram and pulse profile for \mbox{PSR J2111+4606}.
	 The plots have identical form as those shown in Figure~\ref{f:newpulsar-J0106}.}
\end{figure}

\begin{figure}
\centering
	\hspace{-0.2cm}
		\subfigure
		{\includegraphics[width=0.49\columnwidth]{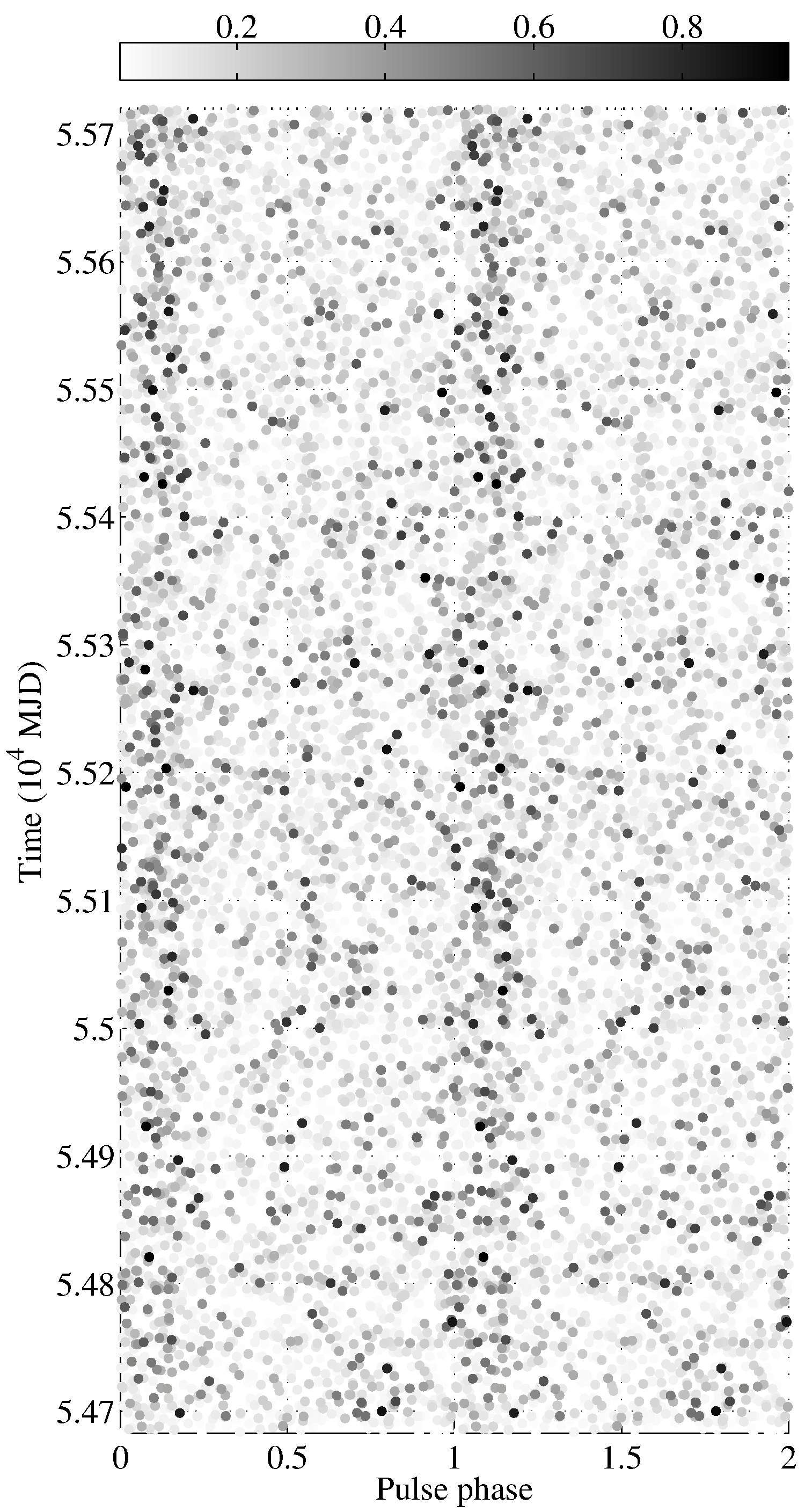}}\hspace{0.1cm}
		\subfigure
		{\includegraphics[width=0.49\columnwidth]{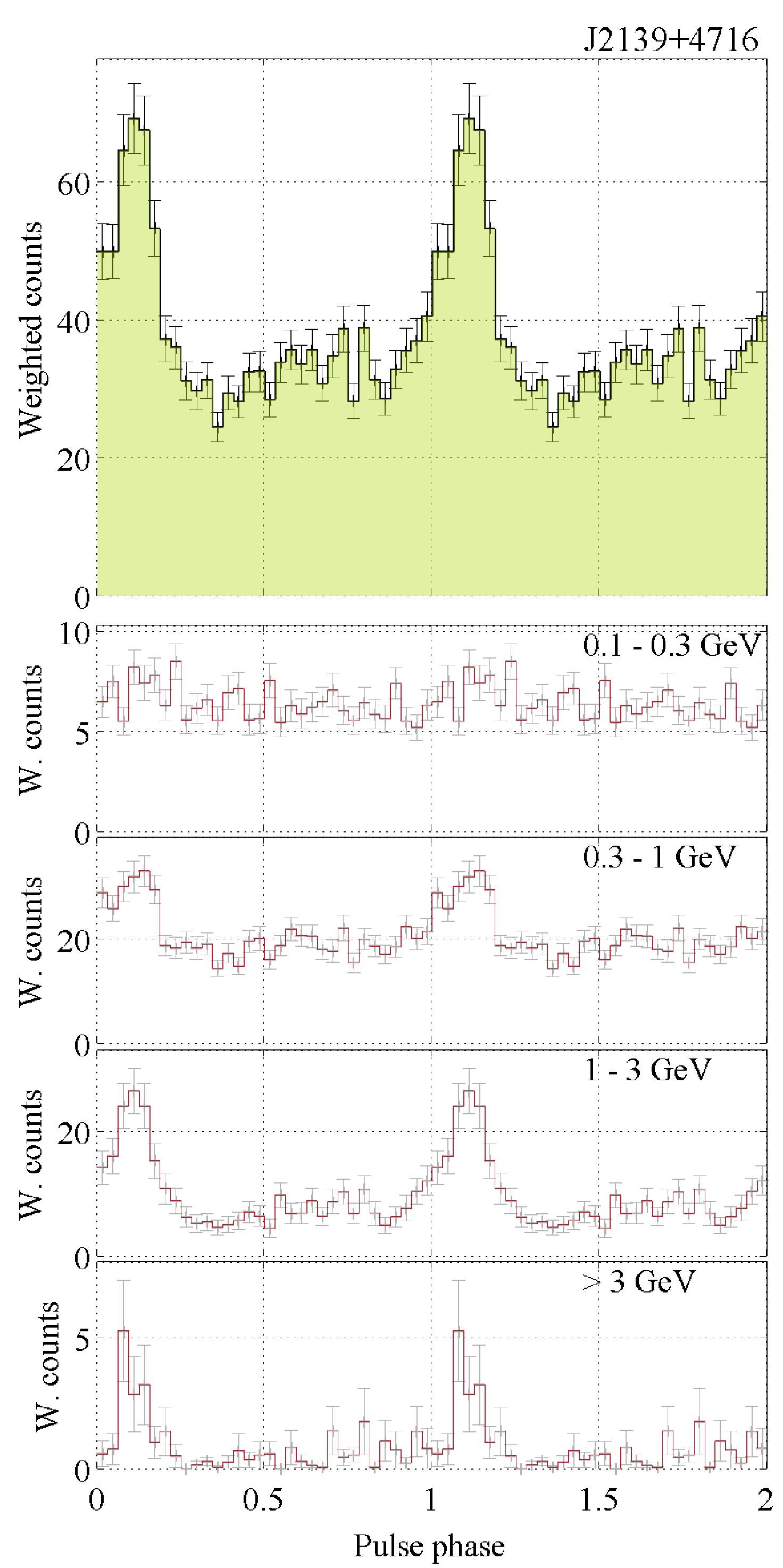}}
	\caption{\label{f:newpulsar-J2139} 
	 Phase-time diagram and pulse profile for \mbox{PSR J2139+4716}.
	 The plots have identical form as those shown in Figure~\ref{f:newpulsar-J0106}.}
\end{figure}

Based on these timing solutions, Figures~\ref{f:newpulsar-J0106}--\ref{f:newpulsar-J2139} 
show the resulting phase-time diagrams and pulse profiles for each of the
newly-discovered pulsars.  These plots are obtained from calculating the
phase for each of the 8000 photons selected for the blind search and using
the parameters listed in Tables~\ref{t:pulnames}
and~\ref{t:pulspinparams}, and weighting each event by its probability
of originating from the pulsar~$w_j$.  
The integrated pulse profiles (weighted pulse phase histograms)
are constructed with a resolution of 32 bins in phase per rotation.
The 1$\sigma$ error bars in the integrated pulse profiles 
are statistical only and are given by $(\sum_j w_j^2)^{1/2}$, 
where $j$ runs over all events falling into the same phase bin. 
(Note that the formula for the \emph{fractional} error has
the \emph{opposite} sign of the exponential: -1/2.)

As seen in Figures~\ref{f:newpulsar-J0106}--\ref{f:newpulsar-J2139},
eight of the nine pulsars have pulse profiles with two peaks.  For
pulsars where the two peaks are separated by nearly one-half of a
rotation it is possible to detect or discover the pulsar at the second
harmonic (i.e. at twice the actual spin frequency of the pulsar).  For
pulsar candidates which were discovered with (apparently)
single-peaked profiles, we tested for the true fundamental spin
frequency by folding at the subharmonic of the discovery frequency. If
the subharmonic is the correct frequency, the two resulting peaks may
satisfy one or more of the following conditions: offsets that are
measurably different than~0.5 in phase, significant differences in the
integrated weighted counts under each peak, or significant differences
in the shape of the peaks in different energy bands.  With modest
signal-to-noise ratios, such determinations are not always conclusive.  For
\mbox{PSR J0106+4855}, our identification of the true period was
subsequently confirmed by the detection of radio pulsations with a
single peaked profile.  For the one apparently single-peaked pulsar in
our sample, \mbox{PSR J2139+4716}, none of the above tests yield
strong evidence for the profile being double peaked at half the
frequency.  Additional data will be required to strengthen this
conclusion.

In order to further characterize each pulse profile, we fit the
pulsars' weighted gamma-ray peaks to Lorentzian lines. The derived
pulse shape parameters are listed in Table~\ref{t:lcparams}.  Note
that \mbox{PSRs J0622+3749} and J1746--3239 show indications of
sub-structures in their main gamma-ray peak. For these pulsars a
single Lorentzian line function is used for fitting the main
component.  Apart from \mbox{PSR J0106+4855}, which is detected at radio
wavelengths (cf. Figure~\ref{f:radioJ0106}), 
there is no particular reference for absolute phase of these pulsars.
For the other eight pulsars we have arbitrarily assigned the absolute
phase reference such that the first peak occurs at a value of~0.1 in phase.
The gamma-ray pulse profiles shown in Figures~\ref{f:newpulsar-J0106}--\ref{f:newpulsar-J2139} 
along with the pulse-profile parameters listed in Table~\ref{t:lcparams} are very
similar to those of the previously-discovered gamma-ray pulsars in
blind searches \citep{16gammapuls2009,8gammapuls2010,FermiPSRCatalog},
further supporting the theory that the gamma-ray emission consists of
fan beams produced in the outer magnetosphere.

\subsection{Spectral parameters}

The spectral parameters for the new pulsars are obtained by
fitting each phase-averaged spectrum with an exponentially cut-off power-law 
with a photon index~$\Gamma$ and a cutoff energy~$E_c$. 
The results for each pulsar are listed in Table~\ref{t:specparams}.
In addition to~$\Gamma$ and~$E_c$ which are explicit parameters of the fit, 
Table~\ref{t:specparams} also gives the important derived physical quantities 
of photon flux~$F_{100}$ (in units of photons cm$^{-2}$ s$^{-1}$) and 
the energy flux $G_{100}$ (in units of erg cm$^{-2}$ s$^{-1}$)
for events with energies between 100~MeV and 100~GeV.

Analogous to the pulse-profile properties, $\Gamma$ and~$E_c$ 
measured for the new pulsars are also similar to those observed for 
previously-detected gamma-ray pulsars \citep{FermiPSRCatalog}. 
This is not surprising, because (as described in Section~\ref{s:dataselection})
target sources for our search have been selected based on similarity of their
spectral properties to known gamma-ray pulsars.

The distance (3~kpc) for one of the new pulsars (PSR~J0106+4855) can be
inferred based on the dispersion of the radio pulse
measuring the free electron column density (see Section~\ref{ss:pulsedradio} for details).
As the remaining eight pulsars are radio-quiet, this method
cannot be used to estimate their distance. 
Furthermore, none of the pulsars is associated with
a known supernova remnant, preventing us from deriving distance
estimates from such source associations.

However, it is still possible to obtain a crude estimate of the distance to 
the new pulsars, by exploiting the observed correlation between the 
gamma-ray luminosity~$L_\gamma$ and the spin-down power~$\dot E$ for other 
gamma-ray pulsars with distance measures \citep[cf.][]{FermiPSRCatalog}.
Based on this correlation ``pseudo gamma-ray luminosities'' $L_{\rm ps}$
are derived as
\begin{equation}
   L_{\rm ps} \sim 3.2 \times 10^{33} ({\dot E / 10^{34}\, \rm{erg \ s}^{-1} })^{1/2}\, \rm{erg \ s}^{-1},
\end{equation}
where the $\dot E$ values are obtained from Table~\ref{t:pulspinparams}. 
Assuming a geometrical correction factor~$f_\Omega=1$ for the emission cone \citep{Watters2009}
for all gamma-ray pulsars, the relation \mbox{$L_\gamma = 4 \pi f_\Omega\, G_{100}\, d^2$}
is used to convert the energy flux and pseudo gamma-ray luminosity
into a ``pseudo distance'' $d_{\rm ps}$, following Equation~(2) of \citet{8gammapuls2010}:
\begin{equation}
    d_{\rm ps} \sim  1.6 \, \frac{\left( {\dot E / 10^{34}\, \rm{erg \ s}^{-1} } \right)^{1/4}}{\left(G_{100} / 10^{-11}\ \rm{erg} \ \rm{cm}^{-2} \ \rm{s}^{-1}\right)^{1/2}} \, {\rm kpc}.
\end{equation}
 For each of the new pulsars the resulting values for $L_{\rm ps}$ and $d_{\rm ps}$ are  
shown in Table~\ref{t:specparams}.
Note that these estimated gamma-ray luminosities and distances are subject
to a number of caveats, detailed in \citet{8gammapuls2010}, and could differ \emph{significantly} 
from the actual values.

\subsection{Why were the new pulsars not found in previous blind searches?}

To examine whether the nine pulsars found with this new method could
be detected with previous methods, we apply
the same search method \citep{TimeDiffTech2006,Ziegler2008} used
to successfully discover the 26 previously-found blind-search LAT
gamma-ray pulsars \citep{16gammapuls2009,8gammapuls2010,2gammapuls2011}.  
We select input data as done in previous searches; events are selected based on
a fixed ROI and minimum-energy cut as described in \cite{8gammapuls2010}.  
We use the same coherence window size ($T=2^{19}$s) as in the previous 
(and in the first stage of this paper's search ) search. 
No photon weights are computed or used.  No sky-gridding
is done in the first stage of the search: only the 2FGL-catalog sky
position is used.

The previous blind search code recovers three of
the nine pulsars (PSRs J1620--4927, J2028+3332 and J2111+4716)
when the 2FGL sky locations are searched.  
Two of the remaining six pulsars are detected \emph{only} if the correct
known sky position is used (as opposed to the 2FGL-catalog
position). If in addition the ROI and energy cuts are optimized
(scanning different values) then three of the remaining four pulsars
are detected.  Finally, if the coherence window size is doubled (which
dramatically raises the computational burden in a full blind search)
then the last pulsar is found.

As compared to the previously-published blind-pulsar searches of LAT data, 
the methods used for this work incorporate several significant improvements 
in sensitivity and computational efficiency, as well as a longer data set, 
that explain why these new discoveries are made.  
First, the use of efficient parameter space gridding over both sky position 
and frequency derivative allows pulsars to be found that are much farther from the
LAT catalog position than is possible with previous searches.  
In addition, using photon weights for both event selection and for the
search computations ensures that the detection significance is near
optimal with only a single trial over event selections.  In contrast,
methods that use a ``cookie cutter'' event selection must either search
over two additional parameters (minimum energy and radius for the
selection), or suffer a sensitivity penalty from potentially
non-optimal event selections.  A key factor is that while weighting the
photons provides only a modest sensitivity improvement over
\textit{optimal} cookie cutter selections, the improvement can be
large when compared to non-optimal cuts.  

One might think that the additional systems found in this paper come
about purely from the significantly-increased computer power that was
available.  This is not the case: the improved methods deserve almost
all of the credit. The improved methods are so computationally
efficient that had we searched only up to 64~Hz as was done in
previous searches, we could have searched all 109 selected 2FGL sources 
over the initial spin-down range (\mbox{$|\dot f| \le$
  5$\times$10$^{-13}$ Hz s$^{-1}$}) and would have found seven of the
nine new pulsars using less than 5000 CPU~hours.  These are modest
resources in comparison with those used in previous blind
searches\footnote{The UCSC group has about $30\,000$ CPU~hours/month
  available; their previous search of 2~years of \emph{Fermi}-LAT data
  used $65\, 000$ CPU~hours.}.  
The remaining two pulsars would only have been found if the spin-down range 
were increased by a factor of twenty to \mbox{$|\dot f| \le$ $10^{-11}$ Hz s$^{-1}$},
increasing the required CPU time to about $100\,000$~hours.

\section{Radio and X-ray Counterpart Searches}
\label{s:counterparts}

\subsection{Pulsed radio emission}
\label{ss:pulsedradio}

These discoveries represent a substantial increase in the number of
gamma-ray pulsars detected in blind searches of \emph{Fermi}-LAT data.
Of the~26 previously-discovered LAT blind-search pulsars, only~3
have been found to pulse in the radio band
\citep{Camilo2009,FermiJ1907+0602,8gammapuls2010,2gammapuls2011} 
and there are tight upper limits on the others \citep{Ray2011,Keith2011}. 
In order to exploit these new discoveries in population studies of the relative beaming 
fraction and geometry of the radio and gamma-ray emission, it is essential to determine 
if they are also visible as radio pulsars.

Because the source list was chosen from \emph{Fermi}-LAT pulsar-like unassociated
sources, all of these sources have been previously searched for radio
pulsations by the \emph{Fermi} Pulsar Search Consortium
\citep{Ransom2011,Hessels2011}.
We have re-analyzed all of these archival observations, with increased
sensitivity because we can now do a single frequency trial folding the
radio data with the gamma-ray ephemeris, and search only in a single
parameter, the dispersion measure (DM). This greatly reduces the number of
points searched in parameter space (the ``trials factor'') and implies
that much smaller pulsed signal amplitudes are statistically
significant.  Where we saw an opportunity to go significantly deeper,
we also made a number of new radio observations.

The telescopes and observing configurations used are described in
Table~\ref{t:RadioObsCodes}, and the characteristics of the individual
radio observations are shown in Table~\ref{t:RadioObsRes}.
We compute the sensitivities using the modified
radiometer equation given in Equation~(A1.22) of \cite{psrhandbook}:
\begin{equation}
  \mathcal{S}_\mathrm{min} = \beta \frac{ 5\, T_\mathrm{sys}}{G \sqrt{n_\mathrm{p} t_\mathrm{int} \Delta F}}     
  \sqrt{\frac{W}{1/f-W}}
  \label{e:smin}
\end{equation}
where $\beta$ is the instrument-dependent factor due to digitization
and other effects (when unknown, we assume $\beta = 1.25$), 
a value of~5 has been assumed for the threshold signal-to-noise ratio
for a confident pulsar detection,
$T_\mathrm{sys} = T_\mathrm{rec} + T_\mathrm{sky}$, $G$ is the
telescope gain, $n_\mathrm{p}$ is the number of polarizations used 
(2 in all cases), $t_\mathrm{int}$ is the integration time, $\Delta F$ is
the observation bandwidth, $f$ is the pulsar spin frequency, and $W$
is the pulse width (for uniformity, we assume $W=0.1/f$).  
Note that conventionally the 3K background temperature is included 
in the receiver temperature $T_\mathrm{rec}$, which is
measured on cold sky, and so $T_\mathrm{sky}$ represents the excess 
temperature from the Galactic synchrotron component, which we estimate 
by scaling the 408~MHz map of \citet{Haslam1982} to the 
observing frequency~$\nu$ with a spectral index $\alpha = -$2.6
(defined as $\mathcal{S}_\nu \propto \nu^{\alpha}$).

We use a simple approximation of a telescope beam response to adjust
the flux sensitivity in cases where the pointing direction was offset
from the true direction to the pulsar.  This factor is given by
\mbox{$q = \rme^{ -(\theta/\mathrm{HWHM})^2/1.5 }$},
where $\theta$ is the offset from the beam center and HWHM is the beam
half-width at half maximum.  A computed flux density limit of
$\mathcal{S}_\mathrm{min}$ at the beam center is thus corrected via
division by~$q$ for targets offset from the pointing direction.

For eight of the nine pulsars, we have established that they are indeed 
radio-quiet (or extremely radio-faint), as viewed from Earth.

\begin{figure}
  \includegraphics[width=\columnwidth]{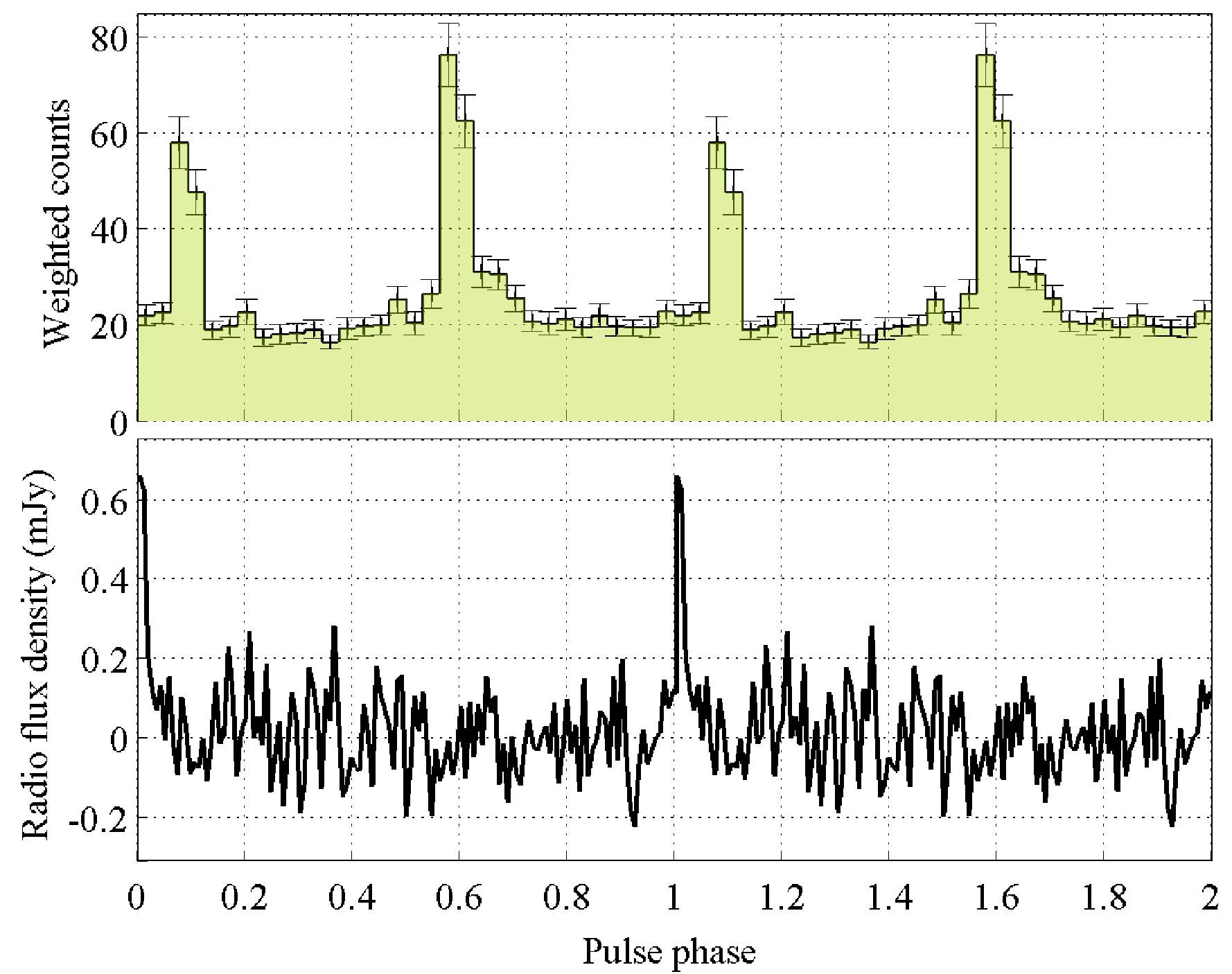}
   \caption{Phase-aligned gamma-ray (top) and radio (bottom) pulse profiles 
   for \mbox{PSR J0106+4855}.\label{f:radioJ0106}}
\end{figure}

For \mbox{PSR J0106+4855} our observations and analysis have revealed very faint radio pulsations. 
In two of our archival 45-minute GBT observations at 820~MHz, we detect the pulsar
at a DM of \mbox{70.87 $\pm$ 0.2 pc cm$^{-3}$} with a single-trial significance of 4--5$\sigma$ in
each observation.  Although the detections are not individually very strong, we gain additional
confidence in their veracity from the fact that the peak phases are consistent to 0.01 pulse
periods when each observation is folded using the ephemeris determined from our gamma-ray timing.
The summed radio profile shows a narrow peak with a duty cycle of \mbox{$\sim$ 2\%},
as shown in Figure~\ref{f:radioJ0106}. We estimate a flux density of \mbox{$\sim$ 20 $\mu$Jy} 
in both observations, using the standard radiometer equation and a measurement of the 
off-pulse noise level. 
These radio pulsations are detected at a flux density below the nominal detection limit 
of 30~$\mu$Jy because the duty cycle is a factor of 5  smaller than the 10\% duty cycle used 
in the sensitivity calculation for unknown pulse shapes.
This corresponds to an equivalent flux density of 8~$\mu$Jy at 1400~MHz using a
typical pulsar spectral index $\alpha = -$1.7.
As seen in Table~\ref{t:RadioObsRes}, we have made two other observations of this source, 
one at 350~MHz with the Green Bank Telescope and one at 1.4~GHz with Effelsberg.  
Neither of these observations detect the radio pulsations. Accounting for
this narrow duty cycle of the pulse, the minimum detectable flux density for the Effelsberg observation 
at 1.4~GHz was about 15~$\mu$Jy, so the non-detection in that observation is not surprising.  
This constrains the spectral index to be $\alpha \le -0.5$.
On the other hand, a spectral index of $-$1.7 would imply
a flux density at 350~MHz of 85~$\mu$Jy, which is above the nominal sensitivity of our 
350~MHz observation. This implies that either $\alpha > -$1.3 or the sensitivity of that observation 
was affected by scatter or DM broadening or higher than expected sky background.
Using the NE2001 model \citep{NE2001}, the measured DM corresponds to an estimated 
distance of 3.0~kpc, over a factor of 2 larger than the
pseudo-distance. Given the radio detection, we can measure the so-called phase lag 
\mbox{$\delta$ $=$ 0.073 $\pm$ 0.003} between the gamma-ray and radio emission.

\subsection{Continuum radio emission}

Pulsars and SNRs (Supernova Remnants) have the same origin,
although the 
comparatively short lifetimes of SNRs means that the
number of pulsar-SNR associations is quite
small. However the rare associations
are of high interest, as they constrain a number of
pulsar and SNR parameters. A recent text-book example is the
association of the gamma-ray pulsar \mbox{J0007+7303} with the shell-type
\mbox{SNR CTA 1} \citep{FermiCTA1,Sun2011}.

For the nine new gamma-ray pulsars no association with a known SNR
listed in the most recent SNR-catalog \citep{GreenSNRcat} is found.
However, SNRs are difficult to identify in case they
are faint, confused or distorted by interaction with dense clouds.
We have used large-scale and Galactic-plane radio continuum surveys
to search for structures in the vicinity of the new pulsars, which may
indicate an association.

For the two pulsars \mbox{PSRs J0106+4855} and \mbox{J0622+3749} located
above 10$\degr$ of Galactic latitude we used the MPIfR-survey 
sampler\footnote{\url{http://www.mpifr.de/survey.html}} 
to extract maps from the 408~MHz and 1420~MHz surveys \citep{Haslam1982,Reich1982}, 
which show faint sources and extended smooth
diffuse emission, but no discrete object. The high-resolution
interferometric NVSS maps \citep{Condon1998} show numerous compact
sources, but no extended features.

The remaining pulsars are located within 4$\degr$ of Galactic latitude
covered by Galactic-plane surveys. The southern-sky pulsar \mbox{PSR J1620--4927}
is located towards the gradient of an extended emission complex as seen in
southern-sky surveys \citep{Reich2001,Jonas1998}. The Southern
Galactic Plane Survey \citep[SGPS,][]{Haverkorn2006} is insensitive to
large scale emission, but shows no small scale structures within 0$\fdg$5
of the pulsar. \mbox{PSR J1746--3239} is located at $b = -2\fdg2$ close to a
emission ridge sticking out from the plane. The shell-type \mbox{SNR G356.6--1.5}
with a size of $20\arcmin \times 15\arcmin$ \citep{Gray1994} is part
of this ridge and about 0$\fdg$6 away from the pulsar. \mbox{PSR J1803--2149}
is located just 2$\arcmin$ away from the peak of a flat spectrum 6~Jy
thermal source visible in all Galactic-plane surveys. \citet{Quireza2006}
list details for the HII-region with a deconvolved Gaussian size of
2$\farcm$7 and a distance of 3.4~kpc. Also CO-emission was 
observed \citep{Scoville1987}. The pulsar is located at the periphery 
of this emission complex, but further studies are needed to investigate this region.
\mbox{PSR J2028+3332} is located at the southern boundary of the thermal Cygnus-X
complex. The 11~cm survey \citep{Reich1984} shows patchy structures in
the field and an extragalactic 1.8~Jy source 40$\arcmin$ distant from
the pulsar. \mbox{PSR J2030+4415} is seen towards the northern periphery of 
strong complex emission from the Cygnus-X region. Dedicated studies are needed 
to find any emission associated with the pulsar. \mbox{PSRs J2111+4606} and 
\mbox{J2139+4716} are both located in low emission areas, where the 21~cm and
11~cm Effelsberg Galactic-plane surveys show faint structures close to the noise level.

\subsection{X-ray}

Subsequent to the pulsar discoveries, we searched for archival X-ray
observations covering the new pulsars' sky locations.  As listed in
Table~\ref{t:xray}, we have found short {\it Swift} observations
(3--10~ks exposure) for five of the pulsars. In addition there is a
6~ks-long {\it XMM-Newton} observation for PSR~J1620--4927 and a
23~ks-long {\it Suzaku} observation for PSR~J0106+4855.  For the two
remaining pulsars, PSRs~J1746--3239 and J2028+3332, 10~ks-long 
{\it Swift} observations were carried out following the discoveries.

These X-ray data were analyzed, and no X-ray counterparts were found for
any of the new gamma-ray pulsars.

In order to estimate upper limits on the X-ray flux for each new
pulsar, a power-law spectrum with a photon index of~2 and a
signal-to-noise of~3 is used.  The values for the absorbing columns
are estimated analogously to \cite{Marelli2011}.  In the 0.3--10~keV
energy range, the derived upper limits on the X-ray flux of the pulsars
are \mbox{between 1 and 3$\times$10$^{-13}$ erg cm$^{-2}$s$^{-1}$}, 
except for PSR~J1620--4927, where an upper-limit
X-ray flux of \mbox{6$\times$10$^{-14}$ erg cm$^{-2}$s$^{-1}$} is
obtained, and PSR~J0106+4855, with an upper-limit X-ray flux
of \mbox{2.3$\times$10$^{-14}$ erg cm$^{-2}$s$^{-1}$}.
These results, listed in Table~\ref{t:xray}, are consistent
with the gamma-to-X-ray flux-ratios for previously-found
\emph{Fermi}-LAT pulsars \citep{Marelli2011}.

\section{Conclusion}

This work reports on the discovery of nine gamma-ray pulsars through
the application of a new blind-search method to about 975~days of
\textit{Fermi}-LAT data. The new pulsars were found by searching
unassociated sources with typical pulsar-like properties selected from
the \textit{Fermi}-LAT Second Source Catalog
\citep{FermiSecondSourceCatalog}.

The sensitivity of blind searches for previously unknown gamma-ray
pulsars is limited by finite computational resources. 
Thus efficient search strategies are necessary maximizing the overall 
search sensitivity at fixed computing cost. 
We have developed a novel hierarchical search method, adapted from 
an optimal semi-coherent method \citep{PletschAllen} together with a sliding
coherence window technique \citep{PletschSLCW2011} originally
developed for detection of continuous gravitational-wave signals from
rapidly-spinning isolated neutron stars. The first stage of the method
is semi-coherent, because coherent power computed using a window of
6~days is incoherently combined by sliding the window over the entire
data set.  In a subsequent stage, significant semi-coherent candidates
are automatically followed up via a more sensitive fully-coherent
analysis.  The method extends the pioneering methods first described
in \citet{TimeDiffTech2006}.

The new method is designed to find isolated pulsars up to~kHz spin
frequencies, by scanning a template grid in the four-dimensional
parameter space of frequency, frequency-derivative and sky location.
The optimal and adaptive sky gridding (not done in
previously-published blind searches) is necessary, particularly at the
higher search frequencies, because the source-catalog sky-positions
are not precise 
enough to retain most of the signal-to-noise ratio in year-long data sets.
A fundamental new element of the method is the exploitation of
a parameter-space metric (well-studied in the continuous
gravitational-wave context \citep[see e.g.][]{bc2:2000,prix:2007ks})
to build an efficient template grid, as well as a metric approach to
construct an optimal semi-coherent combination step
\citep{PletschAllen,pletsch:scmetric}.  A further enhancement over
previous searches is the sub-division of the total search frequency
range into bands via complex heterodyning.  This accommodates memory
limitations, parallelizes the computational work, and permits the use
of efficient sky grids adapted to the highest frequency searched in
each band.  A photon probability weighting scheme
\citep{KerrWeightedH2011} is also used for the first time in a published blind
search, further improving the search sensitivity.

The nine newly-discovered pulsars increase the previously-known
\textit{Fermi}-LAT blind-search pulsar population
\citep{16gammapuls2009,8gammapuls2010,2gammapuls2011} by more than
one-third, and brings the total number to~35.  The inferred parameters
of the new pulsars suggest that they belong to the same general
population as the previously-found blind-search gamma-ray pulsars.
Deep follow-up observations with radio telescopes have been conducted
for all of the new pulsars, but significant radio pulsations have only
been found for \mbox{PSR J0106+4855}.  The null results for the other eight
pulsars indicate that they belong to the growing population of
radio-quiet gamma-ray pulsars, which can only be detected via their
gamma-ray pulsations.
 
The computational work of the search has been done on the
6720-CPU-core \textit{Atlas} Computing Cluster \citep{AtlasArticleMPG}
at the Albert Einstein Institute in Hannover. Recently, in August
2011, we moved the computational burden of the search onto the
volunteer distributed computing system \EAH\footnote{\EAH\ is
  available at \textrm{http://einstein.phys.uwm.edu/}.}
\citep{S4EAH,S5R1EAH,EahRadiopulsar1}. This will provide significantly
more computing power, and will allow a complete search of the
parameter-space up to kHz pulsar spin frequencies.  We also hope that
in the future an improved version of these methods can be used to
carry out blind searches for gamma-ray pulsars in binary systems.

The combination of improved search techniques and much more powerful
computational resources leave us optimistic that we can find still
more gamma-ray pulsars in the \textit{Fermi}-LAT data.  These advances
should also greatly increase the chance of finding the first
radio-quiet gamma-ray MSP with the \mbox{\textit{Fermi}-LAT}.  We hope
that further discoveries, and further study of these systems, will
eventually provide important advances in our understanding of pulsars,
and of their emission mechanisms and geometry.

\acknowledgements

We thank Mallory Roberts, Ramesh Karuppusamy, Joris Verbiest, and
Kejia Lee, for help retrieving and analyzing archival radio data, and
for comments on and corrections to the manuscript, and to J. Eric
Grove for similar assistance with archival X-ray data.  We are
grateful to Robert P. Johnson for his helpful comments on the
manuscript, and to David Thompson for checking the EGRET 3EG catalogs,
and for his management and organizational efforts on our behalf.

This work was partly supported by the Max Planck Gesellschaft and by
U.S. National Science Foundation Grants 0555655 and 0970074.

The National Radio Astronomy Observatory is a facility of the National Science 
Foundation operated under cooperative agreement by Associated Universities, Inc.

The \textit{Fermi}-LAT Collaboration acknowledges generous ongoing
support from a number of agencies and institutes that have supported
both the development and the operation of the LAT as well as
scientific data analysis.  These include the National Aeronautics and
Space Administration and the Department of Energy in the United
States, the Commissariat \`a l'Energie Atomique and the Centre
National de la Recherche Scientifique / Institut National de Physique
Nucl\'eaire et de Physique des Particules in France, the Agenzia
Spaziale Italiana and the Istituto Nazionale di Fisica Nucleare in
Italy, the Ministry of Education, Culture, Sports, Science and
Technology (MEXT), High Energy Accelerator Research Organization (KEK)
and Japan Aerospace Exploration Agency (JAXA) in Japan, and the
K.~A.~Wallenberg Foundation, the Swedish Research Council and the
Swedish National Space Board in Sweden.

Additional support for science analysis during the operations phase is
gratefully acknowledged from the Istituto Nazionale di Astrofisica in
Italy and the Centre National d'\'Etudes Spatiales in France.

\begin{deluxetable*}{llllrr}
\tablecolumns{6}
\tablecaption{Names and Sky Locations of the Discovered Gamma-Ray Pulsars\label{t:pulnames}}
\tablehead{\colhead{Pulsar Name} & \colhead{Source Association}  
& \colhead{R.A.\tablenotemark{a}} & \colhead{Decl.\tablenotemark{a}} &   \colhead{$l$\tablenotemark{b}} & \colhead{$b$\tablenotemark{b}}\\
&  & \colhead{(hh:mm:ss.s)} & \colhead{(dd:mm:ss.s)} & \colhead{(deg)} & \colhead{(deg)}}
\startdata
J0106+4855  &  2FGL\,J0106.5+4854      & 01:06:25.06(1) & +48:55:51.8(2) & 125.5 & -13.9 \\
                         &  1FGL\,J0106.7+4853     & & & &   \\ 
J0622+3749  &  2FGL\,J0621.9+3750      & 06:22:10.51(2) & +37:49:13.6(9) & 175.9 & 11.0 \\
                         &  1FGL\,J0622.2+3751      & & & &   \\
J1620--4927  &  2FGL\,J1620.8--4928     & 16:20:41.52(1) & --49:27:37.1(3) & 333.9 & 0.4 \\ 
		      &  1FGL\,J1620.8--4928c    & & & \\
		      &  1AGL\,J1624--4946 	& & & \\
J1746--3239  &  2FGL\,J1746.5--3238      & 17:46:54.947(8) & --32:39:55.8(7) & 357.0 & -2.2 \\
		       &  1FGL\,J1746.7--3233  	& & & \\
J1803--2149  &  2FGL\,J1803.3--2148 	& 18:03:09.632(9) & --21:49:13(4) & 8.1 & 0.2 \\
	   	       &  1FGL\,J1803.1--2147c	& & & \\
		       &  1AGL\,J1805-2143 	& & & \\
J2028+3332   &  2FGL\,J2028.3+3332     & 20:28:19.860(5) & +33:32:04.36(7) & 73.4 & -3.0 \\
                           & 3EG\,J2027+3429         & & & & \\
J2030+4415   &  2FGL\,J2030.7+4417 	& 20:30:51.35(4) & +44:15:38.1(4) & 82.3 & 2.9 \\
	   	       &  1FGL\,J2030.9+4411	& & & \\
J2111+4606   & 2FGL\,J2111.3+4605	& 21:11:24.13(3) & +46:06:31.3(3) & 88.3 & -1.4 \\ 
      	   	       & 1FGL\,J2111.3+4607	& & & \\
       	   	       & 0FGL\,J2110.8+4608 	& & & \\
J2139+4716  &  2FGL\,J2139.8+4714 	& 21:39:55.95(9) & +47:16:13(1) & 92.6 & -4.0 \\ 
      	   	       &  1FGL\,J2139.9+4715	& & & 
\enddata
\tablecomments{
A list of the nine new pulsars reported in this work, showing their sky locations and associations with cataloged gamma-ray sources.
The associations listed include sources from 
the {\it Fermi}-LAT Second Source Catalog~\citep[2FGL,][]{FermiSecondSourceCatalog},
the {\it Fermi}-LAT First Source Catalog~\citep[1FGL, ][]{FermiFirstSourceCatalog}, 
the {\it Fermi}-LAT Bright Source List~\citep[0FGL, ][]{FermiBSL}, 
the {\it AGILE} Catalog~\citep[1AGL, ][]{AGILEcat}, 
and the Third {\it EGRET} Catalog~\citep[3EG, ][]{3EGcat}. }
\tablenotetext{a}{Right ascension (J2000.0) and declination (J2000.0) obtained from the 
timing model, where the numbers in parentheses are the statistical 1$\sigma$ errors in the last digits. }
\tablenotetext{b}{Galactic longitude ($l$) and latitude ($b$), rounded to the nearest tenth of a degree.}
\end{deluxetable*}

 \begin{deluxetable*}{lllrrrrrrr}
\tablecolumns{9}
\tablewidth{0pt}
\tablecaption{Measured and Derived Parameters of the Discovered Gamma-Ray Pulsars  \label{t:pulspinparams}}
\tablehead{\colhead{Pulsar Name} & \colhead{$f$} & \colhead{$\dot{f}$}  & \colhead{Weighted} 
& \colhead{$\tau$} & \colhead{$\dot{E}$} & \colhead{$B_{\textrm{S}}$}  &\colhead{$B_{\textrm{LC}}$}\\
 {} & \colhead{(Hz)} & \colhead{($-10^{-13}$ Hz s$^{-1}$)} & \colhead{$H$-test} 
     & \colhead{(kyr)} & \colhead{($10^{34}$ erg s$^{-1}$)} & \colhead{($10^{12}$\,G)} & {(kG)}}
\startdata
J0106+4855 & 12.02540173638(8)       & \phn0.61881(7) & 843.1 & 3081.1 & 2.9 & 0.2 & 3.0 \\
J0622+3749 & \phn3.00112633651(5) & \phn2.28985(4) & 288.8 & 207.8 & 2.7 & 2.9 & 0.7 \\
J1620--4927 & \phn5.81616320951(5) & \phn3.54782(4) & 566.4 & 259.9 & 8.1 & 1.4 & 2.4 \\
J1746--3239 & \phn5.01149235750(3) & \phn1.64778(3) & 249.8 & 482.2 & 3.3 & 1.2 & 1.3 \\
J1803--2149 & \phn9.4044983174(2)   & 17.25894(6)         & 451.9 & 86.4 & 64.1 & 1.5 & 11.0 \\
J2028+3332 & \phn5.65907208453(2) & \phn1.55563(2) & 1108.3 & 576.8 & 3.5 & 0.9 & 1.5 \\
J2030+4415 & \phn4.4039248637(5)   & \phn1.2576(2)   & 584.8 & 555.2 & 2.2 & 1.2 & 1.0 \\
J2111+4606 & \phn6.3359340865(4)   & 57.4218(3)           & 554.3 & 17.5 & 143.6 & 4.8 & 11.1 \\
J2139+4716 & \phn3.5354509962(2)   & \phn0.2232(2)  & 351.1 & 2511.5 & 0.3 & 0.7 & 0.3
\enddata
\tablecomments{
The reference epoch for all measured rotational parameters is MJD~55225 and the 
time range for all timing models is MJD~54682 -- 55719. 
The derived quantities in columns 5--8 are based on the $f$ and $\dot f$ values
obtained from the timing model and are rounded to the nearest significant digit. 
To model the timing noise present in \mbox{PSR J1803--2149}, a second frequency derivative
is necessary: \mbox{$\ddot{f}=$7.3(8)$\times$10$^{-24}$ Hz s$^{-2}$}.
To model the timing noise present in \mbox{PSR J2030+4415}, higher frequency derivatives 
up to third order are necessary: 
\mbox{$\ddot{f}=-$1.5(3)$\times$10$^{-23}$ Hz s$^{-2}$} and 
\mbox{$\dddot{f}=-$6(2)$\times$10$^{-31}$ Hz s$^{-3}$}.
To model the timing noise present in \mbox{PSR J2111+4606}, higher frequency derivatives 
up to fourth order are necessary: 
\mbox{$\ddot{f}=$2.30(5)$\times$10$^{-22}$ Hz s$^{-2}$},
\mbox{$\dddot{f}=-$7.9(2)$\times$10$^{-30}$ Hz s$^{-3}$} and
\mbox{$\ddddot{f}=$3.2(4)$\times$10$^{-37}$ Hz s$^{-4}$}.
The numbers in parentheses are the statistical 1$\sigma$ errors in the last digits.
}
\end{deluxetable*}

\clearpage

\clearpage

\begin{deluxetable*}{ccccc}
\tablecolumns{5}
\tablecaption{\label{t:lcparams} Pulse-Profile Parameters of the Discovered Gamma-Ray Pulsars} 
\tablewidth{0pt}
\tablehead{\colhead{Pulsar Name} & \colhead{Peak Multiplicity} & \colhead{FWHM$_1$} & \colhead{FWHM$_2$} & \colhead{$\Delta$}} \\
\startdata
J0106+4855 & 2 & 0.02 $\pm$ 0.01 & 0.05 $\pm$ 0.01 & 0.50 $\pm$ 0.01 \\
J0622+3749 & 2 & 0.21 $\pm$ 0.04 & 0.03 $\pm$ 0.03 & 0.47 $\pm$ 0.03 \\
J1620--4927 & 2 & 0.13 $\pm$ 0.04 & 0.15 $\pm$ 0.05 & 0.21 $\pm$ 0.03 \\
J1746--3239 & 2 & 0.32 $\pm$ 0.07 & 0.11 $\pm$ 0.11 & 0.63 $\pm$ 0.06 \\
J1803--2149 & 2 & 0.04 $\pm$ 0.02 & 0.05 $\pm$ 0.03 & 0.40 $\pm$ 0.02 \\
J2028+3332 & 2 & 0.10 $\pm$ 0.02 & 0.04 $\pm$ 0.01 & 0.38 $\pm$ 0.02 \\
J2030+4415 & 2 & 0.09 $\pm$ 0.03 & 0.04 $\pm$ 0.01 & 0.49 $\pm$ 0.02 \\
J2111+4606 & 2 & 0.08 $\pm$ 0.02 & 0.07 $\pm$ 0.02 & 0.31 $\pm$ 0.02 \\
J2139+4716 & 1 & 0.13 $\pm$ 0.03 & \ldots & \ldots \enddata
\tablecomments{
For each of the nine pulsars, we give the parameters describing the shape of the pulse profile,
including the peak multiplicity, the Full-Widths at Half Maxima (FWHM) of the peaks, 
and the separation~$\Delta$ between the gamma-ray peaks 
for pulsars with more than one peak.
}
\end{deluxetable*}

\begin{deluxetable*}{cccllcl}
\tablecolumns{6}
\tablecaption{\label{t:specparams} Spectral Parameters of the Discovered Gamma-Ray Pulsars}
\tablewidth{0pt}
\tablehead{\colhead{Pulsar Name} & \colhead{$\Gamma$} & \colhead{$E_c$} 
 & \colhead{$F_{100}$\tablenotemark{a}} 
 & \colhead{$G_{100}$\tablenotemark{b}} & \colhead{$L_{\rm ps}$} 
 & \colhead{$d_{\rm ps}$} \\
 & & \colhead{(GeV)} & \colhead{($10^{-8}$ photons cm$^{-2}$ s$^{-1}$)} & \colhead{($10^{-11}$ erg cm$^{-2}$ s$^{-1}$)} & \colhead{($10^{33}$ erg s$^{-1}$)} & \colhead{(kpc)} }
\startdata
J0106+4855  & 1.47 $\pm$ 0.23 $\pm$ 0.12 & 3.31 $\pm$ 0.92 $\pm$ 0.08 & \phn2.56 $\pm$ 0.77 $\pm$ 0.32 & \phn2.40 $\pm$ 0.31 $\pm$ 0.04 & \phn5.5 & 1.4\tablenotemark{c}\\
J0622+3749  & 0.59 $\pm$ 0.34 $\pm$ 0.09 & 0.60 $\pm$ 0.13 $\pm$ 0.04 & \phn2.21 $\pm$ 0.35 $\pm$ 0.08 & \phn1.69 $\pm$ 0.15 $\pm$ 0.05 & \phn5.3 & 1.6 \\
J1620--4927 & 1.01 $\pm$ 0.18 $\pm$ 0.05 & 2.44 $\pm$ 0.42 $\pm$ 0.28 & \phn9.61 $\pm$ 1.68 $\pm$ 0.94 & 13.5 $\pm$ 1.0 $\pm$ 1.7 & \phn9.1 & 0.7 \\
J1746--3239 & 1.33 $\pm$ 0.08 $\pm$ 0.35 & 1.65 $\pm$ 0.12 $\pm$ 0.52 & \phn9.97 $\pm$ 0.94 $\pm$ 1.85 & \phn7.86 $\pm$ 0.41 $\pm$ 0.77 & \phn5.8 & 0.8 \\
J1803--2149 & 1.96 $\pm$ 0.11 $\pm$ 0.20 & 5.73 $\pm$ 1.72 $\pm$ 2.07 & 20.7 $\pm$ 3.1 $\pm$ 0.5 & 13.1 $\pm$ 1.1 $\pm$ 2.1 & 25.6 & 1.3 \\
J2028+3332  & 0.86 $\pm$ 0.21 $\pm$ 0.07 & 1.53 $\pm$ 0.24 $\pm$ 0.08 & \phn5.12 $\pm$ 0.87 $\pm$ 0.44 & \phn6.09 $\pm$ 0.41 $\pm$ 0.13 & \phn6.0 & 0.9 \\
J2030+4415  & 1.89 $\pm$ 0.14 $\pm$ 0.22 & 2.16 $\pm$ 0.65 $\pm$ 0.67 & 13.3 $\pm$ 1.4 $\pm$ 0.2 & \phn7.06 $\pm$ 0.48 $\pm$ 0.66 & \phn4.8 & 0.7 \\
J2111+4606  & 1.63 $\pm$ 0.14 $\pm$ 0.05 & 5.43 $\pm$ 1.80 $\pm$ 1.56 & \phn4.39 $\pm$ 0.69 $\pm$ 0.02 & \phn4.13 $\pm$ 0.34 $\pm$ 0.30 & 38.4 & 2.7 \\
J2139+4716  & 0.80 $\pm$ 0.27 $\pm$ 0.02 & 1.02 $\pm$ 0.21 $\pm$ 0.07 & \phn2.65 $\pm$ 0.44 $\pm$ 0.19 & \phn2.51 $\pm$ 0.21 $\pm$ 0.01 & \phn1.8 & 0.8
\enddata
\tablecomments{
This table describes the spectral properties of each of the nine pulsars, modeling each spectrum as 
an exponentially cut-off power-law with photon indices~$\Gamma$ and cutoff energies~$E_c$. 
The spectral parameters listed here for each pulsar are obtained from maximum likelihood fits. 
The first quoted uncertainties are statistical, while the second are systematic and correspond 
to the differences in the best-fit parameters observed when doing the spectral analyses with 
the P6\_V3 IRFs and associated diffuse emission models (namely, the \textit{gll\_iem\_v02} 
map cube and \textit{isotropic\_iem\_v02} template).
For each object, the pseudo gamma-ray luminosity~$L_{\rm ps}$ and the pseudo distance~$d_{\rm ps}$ are
inferred from the apparent spin-down power~$\dot E$ and the energy flux~$G_{100}$ above $100$~MeV.
\emph{Note that these estimated gamma-ray luminosities and distances are subject
to a number of caveats, detailed in \citet{8gammapuls2010}, and could differ significantly from the actual values.}
}
\tablenotetext{a}{Photon flux measured above $100$~MeV.}
\tablenotetext{b}{Energy flux measured above $100$~MeV.}
\tablenotetext{c}{The actual distance is 3.0~kpc, as inferred from the dispersion of the radio pulse
measuring the free electron column density; see Section~\ref{ss:pulsedradio}.}

\end{deluxetable*}

\begin{deluxetable*}{llrrrrrrr}
\tablewidth{0pt}
\tabletypesize{\small}
\tablecaption{Definition of Radio Observing Codes\label{t:RadioObsCodes}}
\tablehead{
\colhead{Obs Code} & \colhead{Telescope} & \colhead{Gain} & \colhead{Frequency} & \colhead{Bandwidth $\Delta F$} & \colhead{$\beta$\tablenotemark{a}} & \colhead{$n_\mathrm{p}$} & \colhead{HWHM} & \colhead{$T_\mathrm{rec}$} \\
 & & (K/Jy) & (MHz) & (MHz) &   &  & (arcmin) & (K)
}
\startdata
GBT-350      & GBT     & 2.0   & 350  & 100 & 1.05 & 2  & 18.5  & 46 \\
GBT-820      & GBT     & 2.0  & 820  & 200 & 1.05 & 2  & 7.9 & 29 \\
GBT-S        & GBT     & 1.9  & 2000 & 700\tablenotemark{b} & 1.05 & 2  & 3.1 & 22 \\
Eff-L1 & Effelsberg & 1.5 & 1400 & 250 & 1.05 & 2 & 9.1 & 22 \\
Eff-L2 & Effelsberg & 1.5 & 1400 & 140 & 1.05 & 2 & 9.1 & 22 \\
Jodrell      & Lovell & 0.9 & 1520 & 200 &  1.05 & 2 & 6.0 & 24 \\
AO-327       & Arecibo & 11   & 327  & 25  & 1.12 & 2  & 6.3 & 116 \\
AO-Lwide     & Arecibo & 10   & 1510 & 300 & 1.12 & 2  & 1.5 & 27 \\
Parkes-BPSR  & Parkes  & 0.735& 1352 & 340 & 1.05 & 2  & 7.0 & 25
\enddata
\tablecomments{The sky locations of all nine pulsars have been searched for pulsating radio emissions.
This Table gives the radio telescope and back-end parameters used in those observations, which are described in Table \ref{t:RadioObsRes}.}
\tablenotetext{a}{Instrument-dependent sensitivity degradation factor.}
\tablenotetext{b}{The instrument records 800 MHz of bandwidth, but to account for a notch filter for RFI and the lower sensitivity near the band edges, we use an effective bandwidth of 700 MHz for the sensitivity calculations.}
\end{deluxetable*}

\begin{deluxetable*}{lllrrrrrr}
\tablewidth{0pt}
\tablecaption{Radio Search Observations of the New Gamma-Ray Pulsars\label{t:RadioObsRes}}
\tablehead{
\colhead{Target} & \colhead{Obs Code} & \colhead{Date} & \colhead{$t_\mathrm{int}$} & 
\colhead{R.A.\tablenotemark{a}} & \colhead{Decl.\tablenotemark{a}} & \colhead{Offset} & \colhead{$T_\mathrm{sky}$} & 
\colhead{$\mathcal{S}_\mathrm{min}$} \\
 & & & {(min)} & {(J2000)} & {(J2000)} & {(arcmin)} & {(K)} & {($\mu$Jy)}
}
\startdata
J0106+4855   & GBT-350     & 2009-10-25 & 32  & 01:06:37.7 & 48:54:11 & 2.7 & 49.3 & 136  \\
             & GBT-820     & 2010-11-17 & 45  & 01:06:35.5 & 48:55:30 & 1.8 & 5.4 & 30\tablenotemark{b} \\
             & GBT-820     & 2010-12-17 & 45  & 01:06:35.5 & 48:55:30 & 1.8 & 5.4 & 30\tablenotemark{b} \\
             & Eff-L2      & 2011-06-01 & 45  & 01:06:25.1 & 48:55:52 & 0.0 & 1.3 & 31 \\
J0622+3749   & GBT-350     & 2009-10-27 & 32  & 06:22:05.5 & 37:51:07 & 2.2 & 46.2 & 131  \\
             & Eff-L1      & 2010-05-14 & 32  & 06:22:14.7 & 37:51:49 & 2.8 & 1.3 & 30   \\
             & Eff-L1      & 2010-02-06 & 10  & 06:22:15.0 & 37:51:48 &  2.8 & 1.3 & 53\\
             & Eff-L1      & 2010-07-10 & 52  & 06:22:13.0 & 37:50:36 & 1.5 & 1.3 & 22 \\
             & Eff-L1      & 2010-07-10 & 55  & 06:22:13.0 & 37:50:36 & 1.5 & 1.3 & 22 \\
             & GBT-820     & 2010-12-12 & 45  & 06:21:59.0 & 37:51:36 & 3.3 & 5.0 & 32 \\
             & GBT-820     & 2010-12-17 & 45  & 06:21:59.0 & 37:51:36 & 3.3 & 5.0 & 32  \\
J1620$-$4927 & Parkes-BPSR & 2009-08-03 & 270 & 16:21:05.5 & -49:30:32 & 4.9 & 16.9 & 42 \\
             & Parkes-BPSR & 2010-11-18 & 144 & 16:20:43.5 & -49:28:24 & 0.9 & 16.9 & 42 \\
             & Parkes-BPSR & 2011-05-10 & 72  & 16:20:41.3 & -49:27:36 & 0.0 & 16.9 & 58 \\
J1746$-$3239 & GBT-S       & 2009-12-23 & 60  & 17:46:47.9 & -32:36:22 & 3.8 & 5.1 & 30 \\
             & GBT-820     & 2010-11-14 & 45  & 17:46:41.0 & -32:36:18 & 4.6 & 51.6 & 85 \\
             & Parkes-BPSR & 2011-05-10 & 72  & 17:46:54.9 & -32:39:55 & 0.0 & 14.1 & 54 \\
J1803$-$2149 & Eff-L1      & 2010-02-13 & 25  & 18:03:12.0 & -21:47:27 & 1.6 & 17.8 & 55\\
             & Eff-L1      & 2010-05-22 & 32  & 18:03:11.7 & -21:47:28 & 1.5 & 17.8 & 48\\
             & GBT-S       & 2010-09-04 & 65  & 18:03:11.7 & -21:47:28 & 1.5  & 7.1 & 13\\
J2028+3332   & GBT-820     & 2009-08-13 & 60  & 20:27:48.0 & 33:32:24 & 6.6 & 10.3 & 46  \\
             & GBT-S       & 2010-09-20 & 30  & 20:28:18.0 & 33:33:23 & 1.4 & 1.0 & 15 \\
             & GBT-820     & 2010-11-22 & 45  & 20:28:19.0 & 33:32:53 & 0.8 & 10.3 & 33 \\
             & GBT-820     & 2010-12-17 & 45  & 20:28:19.0 & 33:32:53 & 0.8 & 10.3 & 33 \\
             & AO-Lwide    & 2011-05-21 & 45  & 20:28:19.9 & 33:32:06 & 0.0 & 2.1 & 4  \\
             & AO-327      & 2011-05-30 & 25  & 20:28:19.9 & 33:32:06 & 0.0 & 112.9 & 142 \\
J2030+4415   & Eff-L1      & 2010-02-07 & 10  & 20:30:55.0 & 44:11:52 & 3.8 & 4.0 & 62 \\
             & Eff-L1      & 2010-05-15 & 32  & 20:30:55.3 & 44:11:53 & 3.8 & 4.0 & 35 \\
             & Eff-L1      & 2010-07-10 & 60  & 20:30:59.2 & 44:15:33 & 1.4 & 4.0 & 23 \\
             & Eff-L1      & 2010-07-30 & 60  & 20:30:59.2 & 44:15:33 & 1.4 & 4.0 & 23 \\
             & GBT-820     & 2010-11-22 & 45  & 20:30:54.7 & 44:16:08 & 0.8 & 16.0 & 38 \\
             & GBT-820     & 2011-05-28 & 183 & 20:30:51.3 & 44:15:38 & 0.0 & 16.0 & 19 \\
             & Eff-L2      & 2011-06-01 & 45 & 20:30:51.5 & 44:15:37 & 0.0 & 4.0 & 35 \\ 
J2111+4606   & GBT-820     & 2009-09-19 & 60 & 21:11:22.8 & 46:05:53 & 0.3 & 16.0 & 33 \\
             & Jodrell     & 2011-06-22\tablenotemark{c} & $14\times 60$ & 21:11:24.0 & 46:06:29 & 0.0 & 9.0 & 14 \\ 
J2139+4716   & Eff-L1      & 2010-07-10 & 60 & 21:39:52.3 & 47:13:43 & 2.6 & 2.0 & 22 \\
             & GBT-350     & 2009-10-25 & 32 & 21:39:53.2 & 47:15:22 & 1.0 & 74.9 & 171\\
             & GBT-820     & 2010-12-11 & 45 & 21:39:53.5 & 47:13:30 & 2.7 & 8.2 & 34  \\
             & GBT-820     & 2010-12-18 & 45 & 21:39:53.5 & 47:13:30 & 2.7 & 8.2 & 34  \\
             & Eff-L1      & 2010-05-15 & 32 & 21:39:56.9 & 47:15:28 & 0.8 & 2.0 & 29 
\enddata
\tablecomments{The sky locations of all nine pulsars have been searched for radio pulsations. 
Only for \mbox{PSR J0106+4855} radio pulsations are detected. The minimum detectable flux density 
$\mathcal{S}_\mathrm{min}$ for each observation is computed at the observing frequency using 
\Eref{e:smin} and the parameters in Table \ref{t:RadioObsCodes}, as described in the text.}
\tablenotetext{a}{Telescope pointing direction (not necessarily source position)}
\tablenotetext{b}{For these two observations, radio pulsations were detected at a flux density
of 20 $\mu$Jy (see text).}
\tablenotetext{c}{Observed 14 times for 1~hour each between this date and 2011-07-11.}
\end{deluxetable*}

\begin{deluxetable*}{llrrrr}
\tablecolumns{6}
\tablecaption{X-ray Coverage of the Discovered Gamma-Ray Pulsars\label{t:xray}}
\tablehead{\colhead{Pulsar Name} & \colhead{Instrument}  
& \colhead{Exposure Time} & \colhead{Absorbing column\tablenotemark{a}}  
&  \colhead{Flux Upper Limit\tablenotemark{b}} & \colhead{$L_{\gamma}/L{_X}$\tablenotemark{c}}\\
&  & \colhead{(ks)} & \colhead{(10$^{21}$ cm$^{-2}$)} & \colhead{(10$^{-13}$ erg cm$^{-2}$ s$^{-1}$)}} & \\
\startdata
J0106+4855 &  \textit{Suzaku}  & 23.0 & 1.0 & 0.843 & $>$285\\
J0622+3749 &  \textit{Swift XRT}  & 4.4 & 1.0 & 2.58 & $>$65\\
J1620--4927 & {\it XMM-Newton} & 6.0 & 4.0 & 0.674 & $>$1988\\ 
J1746--3239 &  \textit{Swift XRT}  & 8.7 & 1.0 & 1.74 & $>$451\\
J1803--2149 &  \textit{Swift XRT}  & 7.7 & 5.0 & 3.26 & $>$34\\
J2028+3332 &  \textit{Swift XRT}  & 10.3 & 1.0 & 1.57 & $>$387\\
J2030+4415 &  \textit{Swift XRT}  & 10.2 & 4.0 & 2.53 & $>$279\\
J2111+4606 &  \textit{Swift XRT}  & 10.1 & 3.0 & 2.25 & $>$183\\ 
J2139+4716 &  \textit{Swift XRT} & 3.2 & 1.0 & 3.20 & $>$78
\enddata

\tablecomments{The sky locations of all nine pulsars have been
  searched for (non-pulsating) X-rays, using both archival
  and new data. No X-ray sources were found at the new pulsar locations, so flux upper limits and lower limits on the gamma-ray to X-ray luminosity ratio are reported.}
\tablenotetext{a}{Estimated analogously to \cite{Marelli2011}.}
\tablenotetext{b}{Upper limit on the unabsorbed flux in the $0.3-10$~keV energy range, 
using an absorbed power-law model with a photon index of~$2$ and a signal-to-noise of~$3$.}
\tablenotetext{c}{Lower limit on the $L_\gamma/L_X$ flux ratio.}
\end{deluxetable*}

\clearpage

\bibliography{ninepulsars} 

\end{document}